\documentclass[a4paper,11pt]{article}
\pdfoutput=1
\usepackage{jcappub}
\usepackage{bbold}
\usepackage{mathtools}
\usepackage{ulem}
\usepackage{slashed} 

\renewcommand{\thesection}{\arabic{section}}

\def\laq{~\raise 0.4ex\hbox{$<$}\kern -0.8em\lower 0.62ex\hbox{$\sim$}~}
\def\gaq{~\raise 0.4ex\hbox{$>$}\kern -0.7em\lower 0.62ex\hbox{$\sim$}~}

\def\beq{\begin{equation}}
\def\eeq{\end{equation}}
\def\bea{\begin{eqnarray}}
\def\eea{\end{eqnarray}}
\def\bean{\begin{eqnarray*}}
\def\eean{\end{eqnarray*}}

\def \pa {\partial}

\def \la {\lambda}

\def \La {\Lambda}

\def \ka {\kappa}

\def \ga {\gamma}

\def \ep {\epsilon}

\def \om {\omega}

\def\laq{~\raise 0.4ex\hbox{$<$}\kern -0.8em\lower 0.62ex\hbox{$\sim$}~}
\def\gaq{~\raise 0.4ex\hbox{$>$}\kern -0.7em\lower 0.62ex\hbox{$\sim$}~}

\def\be{\begin{equation}}
\def\ee{\end{equation}}
\def \ga {\gamma}

\def\beq{\begin{equation}}
\def\eeq{\end{equation}}
\def\bea{\begin{eqnarray}}
\def\eea{\end{eqnarray}}

\def \pa {\partial}

\def \La {\Lambda}

\newcommand{\boe}{{\bf e}}
\newcommand{\bk}{{\bf k}}
\newcommand{\bn}{{\bf n}}

\newcommand{\Ups}{\Upsilon}

\newcommand{\dd}{\partial}
\newcommand{\ds}{{\slash\hspace{-5pt}\dd}}
\newcommand{\bds}{\overline{{\slash\hspace{-5pt}\dd}}}

\newcommand{\Scal}{\mathcal S}

\newcommand{\Tcal}{\mathcal T}
\newcommand{\HH}{\mathcal H}
\newcommand{\Ical}{\mathcal I}

\newcommand{\Lcal}{\mathcal L}
\newcommand{\Vcal}{\mathcal V}
\renewcommand{\Tcal}{\mathcal T}

\def\laq{~\raise 0.4ex\hbox{$<$}\kern -0.8em\lower 0.62ex\hbox{$\sim$}~}
\def\gaq{~\raise 0.4ex\hbox{$>$}\kern -0.7em\lower 0.62ex\hbox{$\sim$}~}

\def\beq{\begin{equation}}
\def\eeq{\end{equation}}
\def\bea{\begin{eqnarray}}
\def\eea{\end{eqnarray}}
\def\bean{\begin{eqnarray*}}
\def\eean{\end{eqnarray*}}

\def \pa {\partial}

\def \la {\lambda}

\def \La {\Lambda}
\def \De {\Delta}
\def \de {\delta}

\def \ka {\kappa}

\def \ga {\gamma}

\def \ep {\epsilon}

\def \om {\omega}

\title{The gauge invariant cosmological Jacobi map from weak lensing at  leading order}

\author[a, b]{Giuseppe~Fanizza,}
\author[b]{Enea~Di~Dio,}
\author[c]{Ruth~Durrer}
\author[d, e]{and Giovanni~Marozzi}

\affiliation[a]{Instituto de Astrof\'isica e Ci\^encias do Espa\c{c}o, Faculdade de Ci\^encias da Universidade de Lisboa, Edificio C8, Campo Grande, P-1740-016, Lisbon, Portugal}
\affiliation[b]{CERN, Theory Department, CH-1211 Geneva 23, Switzerland}
\affiliation[c]{
Universit\'e de Gen\`eve, D\'epartement de Physique Th\'eorique and CAP,
24 quai Ernest-Ansermet, CH-1211 Gen\`eve 4, Switzerland
}
\affiliation[d]{Dipartimento di Fisica, Universit`a di Pisa, Largo B. Pontecorvo 3, 56127 Pisa, Italy and}
\affiliation[e]{Istituto Nazionale di Fisica Nucleare, Sezione di Pisa, Pisa, Italy}

\emailAdd{gfanizza@fc.ul.pt}
\emailAdd{enea.didio@cern.ch}
\emailAdd{Ruth.Durrer@unige.ch}
\emailAdd{giovanni.marozzi@unipi.it}

\abstract{
We compute the weak lensing Jacobi map at first order in perturbation theory and show that it is both, gauge invariant and symmetric. Linear perturbations therefore do not induce any rotation. However, vector and tensor perturbations do induce $B$-modes in the shear. We show that contrary to what is often claimed in the literature, the shear $B$-mode power spectrum is not fully determined by the rotation power spectrum. Also the $E$-mode shear power spectrum is not determined by the convergence power spectrum. While this difference is small for scalar perturbations, it becomes very significant for tensor perturbations, i.e.~gravitational waves.}

\begin{document}

  \begin{minipage}{.45\linewidth}
    \begin{flushleft}
    \end{flushleft}
  \end{minipage}
\begin{minipage}{.45\linewidth}
\begin{flushright}
 {CERN-TH-2022-008}
 \end{flushright}
 \end{minipage}

\maketitle

\section{Introduction}
In this paper we show that the treatment of weak lensing typically found in the cosmology literature is not complete and actually it is not gauge invariant, since often one part of 
 the Jacobi map is neglected. This is unproblematic in a nearly 
 Newtonian situation but becomes relevant on very large 
 scales and, especially, when we include vector and tensor modes.
Working {\it at first order in cosmological perturbation theory}, we obtain the following results when we correctly consider the Jacobi map for lensing:
\begin{itemize}
\item The Jacobi map  is gauge invariant when expressed in terms of $z$ and $\bn$. Here $z$ is the observed redshift and $\bn$ is the observation direction.
\item The Jacobi map is symmetric, hence there is no rotation at first order. We have $\om\equiv 0$ for scalar, vector and tensor perturbations.
\item The $B$-mode of the shear does not vanish for vector and tensor perturbations and, especially, its angular power spectrum is not given by the one of the rotation, $\om$, as claimed in the literature, see e.g.~\cite{Stebbins:1996wx,Hirata:2003ka,Schmidt:2012nw}.
\item The power spectrum of the  convergence $\ka$ and the $E$-mode of the shear are not simply related as claimed in the literature, see e.g.~\cite{Stebbins:1996wx,Hirata:2003ka,Schmidt:2012nw}. We find that while for scalar perturbations this difference is relevant only at large scales, $\ell\lesssim 20$, for tensor perturbations it is significant on all scales.
\end{itemize}

The fact that, at first order in perturbation theory, weak lensing does not induce rotation is already clear by deriving an evolution equation for rotation e.g.~from Eq.~(27) in~\cite{Perlick:2004tq}. This equation is solved fully non-perturbatively by relativistic numerical N-body simulations in~\cite{Lepori:2020ifz}.  Also in the comprehensive work~\cite{Yoo:2018qba} and in the pioneering paper \cite{Yamauchi:2013fra} it is found that rotation vanishes at first order in the observable Jacobi map. The 'simplification' made in the  references~\cite{Stebbins:1996wx,Hirata:2003ka,Schmidt:2012nw} is that they write the two dimensional Jacobi map as
\be\label{e:Jfalse}
D_{ab} =\de_{ab} + \de\theta_{a,b} \,,
\ee
where $\de\theta_a$ is the (2d) coordinate deflection angle on the sphere and hence this matrix has only two degrees of freedom which are cast either in the power spectra of the convergence $\ka$ and the rotation $\om$ or in the $E$- and $B$-modes of the shear. The above coordinate deflection angle may well have a curl in the presence of vector and tensor perturbations, however, as we shall argue in this paper, it is not a measurable quantity and not gauge invariant.
 
In the observable Jacobi map, the deflection angle cannot be expressed in some arbitrary coordinates but it must be defined with respect to a parallel transported Sachs basis of the 'screen'. In Ref.~\cite{DiDio:2019rfy} it is shown explicitly that at first order the rotation of this Sachs  basis exactly cancels the antisymmetric contribution to $D_{ab}$, rendering it symmetric also in the presence of vector and tensor perturbations. This is no longer the case at second order, where the Jacobi map, in general, has four non-vanishing degrees of freedom. But already at the linear level, the spectra of the convergence $\ka$, the rotation $\om$ and the $E$- and $B$-modes of the shear are unrelated.

In the next section we define the Jacobi map non-perturbatively in the geodesic light-cone gauge~\cite{Gasperini:2011us,Fanizza:2013doa,Fanizza:2015swa} and we derive an expression at first order in a generic gauge.
We then show that, when expressed as function of the observed redshift and direction, this map is gauge invariant, and we find that at first order in perturbation theory its rotation vanishes, see also \cite{Grimm:2018nto}\footnote{We note that in this work we neglect the observer terms since they are expected to be irrelevant for multipoles higher than $\ell=2$. The derivation of the linear gauge invariant Jacobi map including the observer terms can be found in \cite{Grimm:2018nto}.}. In Sections~\ref{s:EBmodes} and \ref{s:convergence} we derive expressions for the shear $E$- and $B$-mode spectra as well as for the convergence, and we compare them with the results in the literature. In Section~\ref{s:results} we evaluate the power spectra numerically and in 
 Section~\ref{s:con} we conclude. The details of some lengthy calculations are deferred to several appendices. \vspace{0.5cm}

\noindent {\bf Notation:}\\
We consider a perturbed spatially flat Friedmann-Lema\^itre-Robertson-Walker (FLRW) universe and write the perturbed metric  as
\be\label{e:met1}
ds^2 = a^2(t)\left[ -(1+2\phi)dt^2 - 2B_idx^idt +(\de_{ij} +2H_{ij})dx^idx^j\right] \,,
\ee
where $a$ is the scale factor and $t$ denotes conformal time. The speed of light is $c=1$. 
The perturbations are split into scalar, vector and tensor parts as follows
\be\label{e:met2}
B_i = B^{(v)}_i + B^{(s)}_{,i} 
\ee
with $\dd^i B^{(v)}_i =0$,
and
\bea\label{e:met3}
H_{ij} &=& -\psi\de_{ij} +\left(E_{,ij}-\frac{1}{3}\de_{ij}\De E\right) +H^{(v)}_{i,j}+H^{(v)}_{j,i} + H^{(t)}_{ij} 
\nonumber
\eea
with $\dd^i H^{(v)}_i =0$ and $\dd^i H^{(t)}_{ij} = H^{(t)\,i}_{i} 
=0$.

\section{Weak lensing and the Jacobi map}
\label{s:jacobi}
In Geodesic Light-Cone (GLC) gauge the Jacobi map can be determined exactly, as shown in \cite{Fanizza:2013doa,Fanizza:2014baa}. To recall this derivation, let us  introduce the GLC gauge coordinates which are the proper time in synchronous 
gauge $\tau$, 
a null coordinate $w$, and two angular coordinates $\tilde{\theta}^a$ ($a=1,2$) which specify the negative of the arrival direction of the light ray at the observer. 
The GLC metric is then given by  six arbitrary functions ($\Ups, U^a, \ga_{ab}=\ga_{ba} $),
\beq
ds^2\! = \!\Ups^2 dw^2\!-\!2\Ups dw d\tau+ \! \gamma_{ab}(d\tilde\theta^a\!- \! U^a dw)(d\tilde\theta^b\!-U^b dw)
\,\,\,\,\,\,\,\,\,\,\,,\,\,\,\,\,\,\,\,\,\,\,a,b=1,2
\label{4}
 \eeq
where $\gamma_{ab}$ and its inverse  $\gamma^{ab}$ lower and  raise  two-dimensional indices. 
In  GLC coordinates the past light-cone of a given observer is defined by  $w = w_o =$ constant, and  null geodesics stay at fixed values of the angular coordinates $\tilde{\theta}^a= \tilde{\theta}^a_o =$ constant
(with $\tilde{\theta}^a_o$ specifying the direction of observation).

To clarify the geometric meaning of these variables, let us consider the limiting case  of  vanishing perturbations, a spatially flat FLRW universe with scale factor $a(t)$ and the observer at the coordinate position $r=0$. 
In this case the geodesic light-cone variables are
\bea
&&
w= r+t,\qquad d\tau=a(t)dt,\qquad  \Ups = a(t), \qquad U^a=0,
\nonumber \\ &&
\gamma_{ab}\,d \tilde\theta^a d\tilde\theta^b = a^2(t)\,r^2 (d \theta^2 +\sin^2 \theta d\phi^2)\,.
\label{FR}
\eea

Let us also introduce the so-called Sachs basis  $\left(s_A^\mu\right)$ with $A=1,2$ \cite{Sachs:1961zz,Seitz:1994xf}. For this we consider a light ray with 4-velocity $k^\mu$  in a matter fluid with 4-velocity $u^\mu$. The Sachs basis of the 'screen' normal to both $u^\mu$ and $k^\mu$ is defined by the conditions 
\cite{Fleury:2013sna,Pitrou:2012ge}:
\begin{align}
\label{eq:SachsBasis1}
g_{\mu\nu}s_A^\mu s_B^\nu&=\delta_{AB}  \,,\qquad
s_A^\mu u_\mu =0\,, \qquad 
s_A^\mu k_\mu =0 \,.
\end{align}
Its evolution equation is
\begin{align}
\label{eq:SachsBasis4}
\Pi^\mu_\nu k^\la\nabla_{\la} s_A^\nu &=0 \qquad  \text{with} \quad \Pi^\mu_\nu = \delta^\mu_\nu - \frac{k^\mu k_\nu}{(u^\alpha k_\alpha)^2} - \frac{k^\mu u_\nu + u^\mu k_\nu}{u^\alpha k_\alpha} \, ,
\end{align}
where $\Pi^\mu_\nu$ is a projector on the two-dimensional space orthogonal to $u_\mu$ and to $n_\mu = u_\mu + (u^\alpha k_\alpha)^{-1} k_\mu$ with $n^\alpha n_\alpha = 1$ and $n^\alpha u_\alpha = 0$.
The Sachs vectors $(s_A^\mu)$ together with the photon 4-velocity $k^\mu$ and the matter fluid 4-velocity $u^\mu$ form a 4-dimensional basis $\{u^\mu,k^\mu,s^\mu_1,s^\mu_2 \}$, with 2 spatial vectors, one null and one time-like vector. A possible alternative decomposition, well-suited to describe the geodesic deviation of null geodesics, consists in using a null basis $\{l^\mu,k^\mu,\Sigma^\mu, \bar\Sigma^\mu\}$ for the screen~\cite{Newman:1961qr}, where $l^\mu$ is another null vector normal to the screen such that $l^\mu k_\mu =-1$ and $\Sigma^\mu =\frac{1}{\sqrt{2}}\left( s^\mu_1+i s^\mu_2 \right)$.
This approach has been recently used in ref.~\cite{Ginat:2021nww} in the cosmic ruler framework~\cite{Schmidt:2012ne,Jeong:2013psa}. It is straightforward to verify that the two approaches are equivalent once we set
\beq
l^\mu\equiv-\frac{1}{u^\mu k_\mu} u^\mu -\frac{1}{2\,(u^\mu k_\mu)^2} k^\mu\,.
\eeq
Indeed, in terms of $k^\mu$ and $l^\mu$, we have $l^\mu s^A_\mu = 0$ and the projector $\Pi^\mu_\nu$ in Eq.~\eqref{eq:SachsBasis4} becomes
\beq
\Pi^\mu_\nu
=\delta^\mu_\nu
+k^\mu l_\nu
+l^\mu k_\nu\,.
\eeq

Following \cite{Fanizza:2013doa}, it can be shown that in  GLC coordinates the screen space, normal to incoming photon geodesics and 
the observer's worldline, is simply given by the 2-dimensional subspace spanned by the angles $\tilde\theta^a$. 
In GLC gauge, the angular part of the Sachs basis, which is fixed up to a global rotation, satisfies~\cite{Fanizza:2013doa}
\beq
\gamma_{ab}\,s^a_As^b_B=\delta_{AB}\qquad,\qquad \nabla_\tau s^a_A=0 \, ,
\label{eq:Sachsproperties}
\eeq
and the exact Jacobi map is simply given by~\cite{Fanizza:2013doa,Fanizza:2014baa}
\beq
J^A_B=s^A_a \left( \Delta^{ab} s_{b\,B} \right)_o\,,
\eeq
where $\Delta^{ab}$ is defined by $(\Delta^{ab})=-2(\pa_\tau\gamma_{ab})^{-1}$ and $s_a^A$ are the angular component of the Sachs basis. The GLC line element in Eq.~\eqref{4} still admits a residual gauge freedom for the metric entries allowing for a coordinate transformation which depends only of certain combinations of the coordinates \cite{Fanizza:2013doa,Fleury:2016htl}. This gauge freedom can be fixed to the so-called \textit{observational gauge}. Within this gauge, \cite{Fanizza:2018tzp} in the vicinity of the observer geodesic, $\gamma_{ab}$ can be expanded as $\gamma_{ab}=\hat{\gamma}_{ab}(w-\tau)^2+\mathcal{O}\left((w-\tau)^3\right)$, where $\hat{\gamma}_{ab}=\text{diag}\left( 1,\sin^2\tilde\theta^1 \right)$ is the unperturbed metric of the 2-sphere. With this we obtain
\beq
\pa_\tau \gamma_{ab} = -2 \hat{\gamma}_{ab} \left( w-\tau \right) +\mathcal{O}\left((w-\tau)^2\right)
\eeq
and 
\beq
\Delta^{ab}\approx \hat{\gamma}^{ab}\left( w-\tau \right)^{-1} +\mathcal{O}\left(1\right) \,.
\eeq
We introduce also the Sachs basis of the FLRW metric which is given by $\bar{s}_{a\,A}=\bar{d}\,\hat{s}_{a\,A}$, where $\hat{s}_{a\,A}$ are  2d angular directions normal to the background parts of $u^\mu$ and $k^\mu$ such that $\hat{s}_{a\,A}\hat{s}_{b\,A}=\hat{\gamma}_{ab}$ and $\bar d$ is the (unperturbed) area (or angular diameter) distance in a FLRW universe.
Using also that in the vicinity of the observer worldline the Sachs basis is given by $s_{b\,B}=\hat{s}_{b\,B}\left( w-\tau \right)+\mathcal{O}\left( (w-\tau)^2 \right)$ 
we obtain at the observer position
\bea
\left(\Delta^{ab}s_{b\,B}\right)_o &=& \hat{s}^a_B  \quad \mbox{so that}\\
J^A_B &=&  s^A_a\hat{s}^a_B  \,.
\eea
Until this point, our expressions are  exact. We now continue to first order in perturbation theory where we have\footnote{As explicitly shown in Appendix C of \cite{Marozzi:2016qxl}, Eq.~\eqref{eq:211} is the linear solution of both Eqs.~\eqref{eq:Sachsproperties}, ensuring then that the basis is parallel transported.}
\bea
\label{eq:211}
s_{a\,A} &=& \bar{s}_{a\,A}+\frac{1}{2}\de\gamma_{ab}\bar{s}^b_A \quad \mbox{hence }\\
\label{eq:212}
J^A_B &=&  \bar d\left( \de^A_B  +\frac{1}{2}\de\hat\gamma_{ab}\hat{s}^{bA}\hat{s}^a_B\right)\,.
\eea
Here $\delta\gamma_{ab}\equiv\gamma_{ab}-\bar\gamma_{ab}$ is the linear perturbation of $\gamma_{ab}$ and $\de\hat\gamma_{ab}\equiv\bar{d}^{-2}\de\gamma_{ab}$.
Already from this expression it is clear that the Jacobi map is symmetric at first order in perturbation theory. 
(Note, however that even for a symmetric Jacobi matrix shear can change the direction of the principal axes of an elliptical source, if the principal axes of the source and the Jacobi map are misaligned, see~\cite{Francfort:2021oog} for a detailed study of this effect.)

To complete our calculation we now have to express the first order perturbation of the angular metric in GLC gauge in terms of the general perturbed metric defined in Eqs.~\eqref{e:met1} to \eqref{e:met3}.
This is a lengthy calculation which we perform in Appendix~\ref{sec:A1}. The result is as follows
\bea \label{e:dgaxt}
\delta\gamma^{ab}(t,x) &=& \delta\gamma^{ab}_{GI}+\delta^{kj}\pa_k\bar\gamma^{ab} \chi_j
-2\,\mathcal{H}\,\bar\gamma^{ab}\left( B+E' \right)\,, \\
\mbox{where} && \nonumber\\
\chi_j &=& H_j^{(v)} + E_{,j} \qquad \mbox{and}\\
 \delta\gamma^{ab}_{GI}
&=&\bar{\gamma}^{ac}\pa_c \delta\theta^b_{GI}
+\bar{\gamma}^{bc}\pa_c \delta\theta^a_{GI}
-2\,a^2\,r^2\bar\gamma^{ac}\,H^{(t)}_{cd}\bar\gamma^{db}
+2\,\bar\gamma^{ab}\Psi\,,\\
\delta\theta^a_{GI}
&=& -\int_0^{r_s}dr\,\frac{\hat\gamma^{ac}}{r^2}\left( 
\delta w_{GI\,,c}-r\,\sigma^{(v)}_c-2\,r H^{(t)}_{rc} \right)\,, \label{e:dthetaGI}\\
\delta w_{GI}
&=& -\int_0^{r_s}dr\left( \Phi+\Psi-\sigma_r^{(v)}-H^{(t)}_{rr}\right)\,. \label{e:dwGI}
\eea
Here $\Phi$ and $\Psi$ are the Bardeen potentials and $\sigma^{(v)}_i=B^{(v)}_i+{H^{(v)}_i}'$ is a gauge invariant vector perturbation (see e.g.~\cite{durrer_2020} for an introduction to gauge invariant cosmological perturbation theory). The prime indicates the derivative w.r.t.~conformal time $t$. The integrals in \eqref{e:dthetaGI} and \eqref{e:dwGI} are performed along the unperturbed light path. The quantities indexed by 'GI' are explicitly gauge invariant. Interestingly $\delta\gamma^{ab}(t,x^i)$ is not. The reason for this is that we calculate it at fixed coordinate position and time $(t,x^i)$ which depend on the coordinate system chosen. To convert it to an observable we have to compute it at an observed redshift $z$ and in some fixed observed direction $\bn$. The details of this conversion are given in Appendix~\ref{sec:A2}. The result is
\bea
\delta\gamma^{ab}(z,{\bf n})&=&\bar{\gamma}^{ac} \nabla_c\delta\theta^b_{GI}
+\bar{\gamma}^{bc} \nabla_c\delta\theta^a_{GI}
-2\,a^2\,r^2\bar\gamma^{ac}\,H^{(t)}_{cd}\bar\gamma^{db}
+2\bar\gamma^{ab}\Psi\nonumber \\  &&
+2\bar\gamma^{ab}\left(\frac{\delta z_{GI}}{\mathcal{H}r}
-\delta z_{GI}
+\frac{\delta w_{GI}}{r}\right)\,,\\
\mbox{where} && \nonumber\\
\delta z_{GI} &=& -\Phi
+\sigma^{(v)}_r
+\delta w'_{GI} 
-\frac{1}{a}\pa_r\int_0^{t_s}dt\,a\,\Phi \,.
\eea
The first line is simply $\delta\gamma^{ab}_{GI}$, but in the second line the gauge dependent terms of \eqref{e:dgaxt} have been converted into gauge invariant terms. This 
expression is now explicitly gauge invariant and, thanks to Eq.~\eqref{eq:212}, leads to one of the main result of this work: the  Jacobi map at linear order in perturbation theory is
\bea
J^A_B(z,{\bf n})&=&\bar d(z,{\bf n})
\left\{\delta_{AB}\left[1-\Psi
+\left(1-\frac{1}{\mathcal{H}r}\right)\delta z_{GI}
-\frac{\delta w_{GI}}{r}\right]\right.\nonumber\\
&&\left.-\frac{1}{2}\hat{\gamma}_{ac} \nabla_b\delta\theta^c_{GI}\hat{s}^a_A\hat{s}^b_B
-\frac{1}{2}\hat{\gamma}_{bc} \nabla_a\delta\theta^c_{GI}\hat{s}^a_A\hat{s}^b_B
+H^{(t)}_{ab}\hat{s}^a_A\hat{s}^b_B\right\}\,.
\label{eq:cornerstone}
\eea
This expression  also shows that the Jacobi map is gauge invariant as it is expected for an observable. Furthermore, it is  symmetric which shows that at first order lensing does not induce rotation. The trace of $J_B^A$ gives rise to the convergence $\ka$, while its traceless part defines the shear.

\section{Shear E and B modes}
\label{s:EBmodes}
We follow the standard approach  in the literature, for the decomposition of lensing shear into  its $E$- and $B$- modes. It relies on the definition of an appropriate 2-screen orthogonal to the line-of-sight. We consider the generic line-of-sight
\be
n^i=\left(\sin\theta\cos\phi,\sin\theta\sin\phi,\cos\theta\right)\,
\ee
and define the basis $(e^i_A)$ in the 2d plane orthogonal to $n^i$ through the following conditions
\be
e^i_A e_{iB}=\delta_{AB}
\qquad\text{and}\qquad
e^i_A n_i =0\,.
\ee
The vectors $e^i_A$ can be given explicitly  by
\be
e^i_1=\left( \cos\theta\cos\phi,\cos\theta\sin\phi,-\sin\theta \right)
\qquad\text{and}\qquad
e^i_2=\left( -\sin\phi,\cos\phi,0 \right) \,.
\ee
These are simply the vectors $\hat{e}^i_\theta$ and $\hat{e}^i_\phi$ used in Appendix B of \cite{Schmidt:2012nw}. The standard approach to lensing theory projects $D_{ab}$ as given in \eqref{e:Jfalse} on the angular part of the basis $e^i_A$. Once this projection is done, the decomposition in $E$- and $B$- modes takes place on the 2d screen defined by $e^i_A$. We are going to show here that this procedure is automatically taken into account in Eq.~\eqref{eq:cornerstone} and hence no further projection is needed to extract the shear components.

To this aim, we first recall that
\be
\frac{\pa n^i}{\pa \theta^a}\frac{\pa \theta^a}{\pa n^j}=\delta^i_j-n^i n_j\,.
\ee
Moreover
\be
\hat{s}^a_A=\frac{\pa \theta^a}{\pa n^i}e^i_A
\qquad\text{and}\qquad
e^i_A=\frac{\pa n^i}{\pa \theta^a}\hat{s}^a_A\,.
\label{eq:basis_coord_tr}
\ee
Hence, for a generic spatial tensor $T$ we obtain
\bea
\hat{s}^a_A\hat{s}^b_B\,T_{ab}
&=&\frac{\pa\theta^a}{\pa n^i}\frac{\pa\theta^b}{\pa n^j}\frac{\pa n^k}{\pa\theta^a}\frac{\pa n^l}{\pa\theta^b}e^i_Ae^j_B\,T_{kl}
\nonumber\\
&=&\left( \delta^k_i-n^k n_i \right)\left(\delta^l_j-n^l n_j\right)e^i_Ae^j_B\,T_{kl}
=e^i_Ae^j_B\,T_{ij}\,.
\label{eq:equivalence}
\eea
This proves that Eq.~\eqref{eq:cornerstone} is automatically ready to be used as a projected quantity on the 2d screen for the study of the shear. Here we remark an important point. As already mentioned above, standard approaches to lensing theory \cite{Stebbins:1996wx,Hirata:2003ka,Schmidt:2012nw} start from the amplification matrix $\mathcal{A}_{ij}=\pa_i\delta x_j$ and  project it on the  basis $e^i_A$ afterwards. Following our derivation, this procedure is not well-posed from the mathematical point of view, since $\mathcal{A}_{ij}$ does not necessarily transform as a tensor\footnote{Even though this difference leads only to subleading corrections.} as required instead by Eq.~\eqref{eq:equivalence}. Moreover, in the treatment of weak lensing theory with the Jacobi map the projection on the $e^i_A$ basis naturally emerges.

We now decompose the lensing shear  in $E$- and $B$- modes defined through the circular basis
\be
e^j_\pm\equiv\frac{1}{\sqrt{2}}\left( e^j_1{\pm} i e^j_2 \right)\,,
\ee
where the subscripts $+/-$ respectively refer to anti-clockwise and clockwise circular basis. According to Eqs.~\eqref{eq:basis_coord_tr}, 
\be
e^i_\pm=\frac{\pa n^i}{\pa\theta^a}\hat{s}^a_{\pm}
\label{eq:38}
\ee
where we have defined $\hat{s}^a_\pm\equiv\frac{1}{\sqrt{2}}\left( \hat{s}^a_1{\pm} i \hat{s}^a_2 \right)$. Let us stress for later use that under a rotation by an angle $\beta$ of the basis in the 2d screen, both $e^j_\pm$ and $\hat{s}^a_\pm$ change by an overall phase  $\exp\left(\pm i \beta\right)$.

We first decompose  $J_{AB}(z,{\bf n})$  into
\bea
\kappa(z,{\bf n})\equiv1-\frac{1}{2\,\bar{d}}\,\delta^{AB}J_{AB}
\qquad&&\qquad  \mbox{convergence,} \nonumber \\
\omega(z,{\bf n})\equiv\frac{i}{2\,\bar{d}}\,\sigma_2^{AB}J_{AB}
 \equiv 0  &&\qquad  \mbox{rotation,}\nonumber\\
 \mbox{and shear } \nonumber \\
\ga_1(z,{\bf n})\equiv\frac{1}{2\,\bar{d}}\,\sigma_3^{AB}J_{AB}
\qquad&&,\qquad
\ga_2(z,{\bf n})\equiv\frac{1}{2\,\bar{d}}\,\sigma_1^{AB}J_{AB}\,.
\label{eq:39}
\eea
Here $\left(\sigma_i^{AB}\right)$ are the Pauli matrices. We also introduce
\be
\ga_\pm(z,{\bf n})\equiv\ga_1\pm i\ga_2\,,
\ee
which are the positive and negative helicity  components of $\ga$.
  For scalar perturbations, the fully relativistic and gauge invariant shear agrees with the standard result. Vector perturbations are less interesting as they are not generated in inflationary models. The vector shear from a cosmic string network is estimated in~\cite{Yamauchi:2013fra}.

From \eqref{eq:39}, inserting $J_{AB}$ from \eqref{eq:cornerstone} we find
\be
\ga_\pm\left(z,{\bf n}\right)=
-\hat{\gamma}_{ac} \nabla_b\delta\theta^c_{GI}\hat{s}^a_{\pm}\hat{s}^b_{\pm}
+H^{(t)}_{ab}\hat{s}^a_{\pm}\hat{s}^b_{\pm}\,.
\label{eq:311}
\ee
In this derivation, $\ga_\pm$ naturally emerge as projected onto the circular polarization basis.

Inserting also \eqref{e:dthetaGI} for $\delta\theta^c_{GI}$ and making use of Eq.~\eqref{eq:D7} we obtain
\bea
\gamma_+
&=&\int_0^{r_s}\frac{dr}{r^2}\left( 
\pa_+^2\delta w_{GI}-r\,\pa_+\,\sigma^{(v)}_+-2\,r \pa_+\,H^{(t)}_{r+} \right)
+H^{(t)}_{++}\,,
\nonumber\\
\gamma_-
&=&\int_0^{r_s}\frac{dr}{r^2}\left( 
\pa_-^2\delta w_{GI}-r\,\pa_-\,\sigma^{(v)}_-2\,r \pa_-\,H^{(t)}_{r-} \right)
+H^{(t)}_{-}\,.
\label{eq:shearEB_explicit}
\eea
This result is in agreement with  \cite{Bernardeau:2009bm} where $\pa_\pm$ are written as the more familiar spin raising and spin lowering operators,
 $\pa_+\equiv-\ds/\sqrt{2}$ and $\pa_-\equiv-\bds/\sqrt{2}$, where  (see also Appendix \ref{app:swsh} for details) the helicity components of  $\sigma^{(v)}_a,$ $H_{ra}$ and $H_{ab}$ are  defined by
\be
\sigma^{(v)}_{\pm}\equiv \hat{s}^a_{\pm}\sigma^{(v)}_a
\qquad,\qquad
H^{(t)}_{r\pm}\equiv \hat{s}^a_{\pm} H^{(t)}_{ra}
\qquad\text{and}\qquad
H^{(t)}_{\pm\pm}\equiv \hat{s}^a_{\pm}\hat{s}^b_{\pm} H^{(t)}_{ab}\,.
\label{eq:definitions}
\ee
From Eqs.~\eqref{eq:shearEB_explicit} and \eqref{eq:definitions} it is evident that $\ga_\pm$ are  helicity $\pm 2$ objects. They can now be decomposed in terms of the spin-weighted spherical harmonics $_{\pm 2}  Y_{\ell m}({\bf n})$ (see Appendix \ref{app:swsh} for details) as
\be
\ga_\pm\left(z,{\bf n}\right)=\sum_{\ell m}a^\pm_{\ell m}(z)\,_{\pm 2}Y_{\ell m}({\bf n})\,.
\label{eq:decomposition}
\ee
The coefficients $a^\pm_{\ell m}$ are in general
 complex. They can be further decomposed as
\be
a^E_{\ell m}\left(z\right)\equiv \frac{1}{2}\left( a^+_{\ell m}+a^-_{\ell m} \right)
\qquad\text{and}\qquad
a^B_{\ell m}\left(z\right)\equiv \frac{i}{2}\left( a^+_{\ell m}-a^-_{\ell m} \right)\,.
\label{eq:coeff}
\ee
These are the coefficients of  the $E$- and $B$- mode spectra of the shear, 
\be
C^E_\ell(z_1,z_2)=\langle a^E_{\ell m}(z_1){a^E_{\ell m}}^*(z_2)\rangle
\qquad\text{and}\qquad
C^B_\ell(z_1,z_2)=\langle a^B_{\ell m}(z_1){a^B_{\ell m}}^*(z_2)\rangle\,.
\ee
Here $\langle \dots \rangle$ stands for an {\it ensemble average} over different realizations of the perturbations. Contrary to $\ga_{\pm}$ the shear $E$- and $B$-mode spectra are independent of the chosen coordinate system.  Since they have opposite helicity, $E$- and $B$-modes  are  uncorrelated.

\subsection{Power spectra}\label{s:spectra}
In this section we present the power spectrum for $\ga_\pm(z,{\bf n})$. Inserting \eqref{e:dwGI} in \eqref{eq:shearEB_explicit}  we obtain
\bea
\gamma_\pm(z,{\bf n})
&=&\pa_\pm^2\int_0^{r_s}dr\frac{r_s-r}{r\,r_s}\left( \sigma_r^{(v)}+H^{(t)}_{rr} -\Phi-\Psi\right)
-\pa_\pm\,\int_0^{r_s}\frac{d r}{r}\left( 
\sigma^{(v)}_\pm+2 H^{(t)}_{r+} \right)
+H^{(t)}_{\pm\pm}\,. \nonumber\\
\label{eq:ga_pm}
\eea
We focus our investigation on tensor perturbations. The scalar part is simple and leads to the well known standard result, see, e.g.~\cite{Stebbins:1996wx,Hirata:2003ka,Schmidt:2012nw}, and, as already mentioned, vector perturbations are absent in simple inflationary models. 

In Appendix~\ref{a:shear} we give the detailed derivation of the shear tensor power spectrum using the total angular momentum method. Here we just report the results for the $E$- and $B$-mode spectra. We can decompose each of them in the form
\bea \label{eq:power_spectrum}
C^{X}_\ell(z_1,z_2)&\equiv&\,_0C^{X}(z_1,z_2)+\,_1C^{X}(z_1,z_2)+\,_2C^{X}(z_1,z_2)
\nonumber\\
&&+\,_{01}C^{X}(z_1,z_2)+\,_{02}C^{X}(z_1,z_2)+\,_{12}C^{X}(z_1,z_2)\,, 
\eea
where $0,~1,~2$ respectively label helicity 0, helicity 1  and helicity 2  projections of the tensor modes onto the screen. A single subscript stands for auto-correlation of helicity projections whereas double subscripts indicate cross-correlations between different helicity projections. We obtain the following non-vanishing contributions for the auto-correlations 
\bea
\,_0C^E_\ell(z_1,z_2)
&=&\frac{1}{32\pi}\left[\frac{(\ell+2)!}{(\ell-2)!}\right]^2
\int_0^{r_1}dr\int_0^{r_2}dr'\frac{r_1-r}{r r_1}\frac{r_2-r'}{r' r_2}
\nonumber\\
&&\times
\int k^2dk P_T(k,z,z')
\frac{j_\ell(kr)}{(kr)^2}
\frac{j_\ell(kr')}{(kr')^2}
\nonumber\\
\,_1C^E_\ell(z_1,z_2)
&=&\frac{(\ell-1)^2(\ell+2)^2}{8\pi}
\int_0^{r_1}\frac{dr}{r}\int_0^{r_2}\frac{dr'}{r'}
\int k^2dk P_T(k,z,z')
\nonumber\\
&&\times
\left[ \frac{j_{\ell+1}(kr)}{kr}-(\ell+1)\frac{j_\ell(kr)}{(kr)^2}\right]
\left[ \frac{j_{\ell+1}(kr')}{kr'}-(\ell+1)\frac{j_\ell(kr')}{(kr')^2}\right]
\nonumber\\
\,_2C^E_\ell(z_1,z_2)
&=&\frac{1}{8\pi}
\int k^2dk P_T(k,z_1,z_2)
\left[\frac{\ell(\ell -1)}{2} \frac{j_{\ell }(kr_1)}{(kr_1)^2}
-j_{\ell }(kr_1)+\frac{j_{\ell -1}(kr_1)}{kr_1}\right]
\nonumber\\
&&\times\left[\frac{\ell(\ell -1)}{2} \frac{j_{\ell }(kr_2)}{(kr_2)^2}
-j_{\ell }(kr_2)+\frac{j_{\ell -1}(kr_2)}{kr_2}\right]
\nonumber\\
\,_1C^B_\ell(z_1,z_2)
&=&\frac{1}{8\pi}
\int_0^{r_1}\frac{dr}{r}\int_0^{r_2}\frac{dr'}{r'}
\int k^2dk P_T(k,z,z')
\nonumber\\
&&\times\left[ (\ell^2-\ell-3)\frac{j_\ell (kr)}{kr}+ j_{\ell-1}(kr)+j_{\ell+1}(kr) \right]
\nonumber\\
&&\times\left[ (\ell^2-\ell-3)\frac{j_\ell (kr')}{kr'}+ j_{\ell-1}(kr')+j_{\ell+1}(kr') \right]
\nonumber\\
\,_2C^B_\ell(z_1,z_2)
&=&\frac{1}{8\pi}
\int k^2dk P_T(k,z_1,z_2)
\nonumber\\
&&\times\left[(\ell +2)\frac{ j_{\ell}(kr_1)}{kr_1}-j_{\ell +1}(kr_1)\right]
\left[(\ell +2)\frac{ j_{\ell}(kr_2)}{kr_2}-j_{\ell +1}(kr_2)\right]\,.
\label{eq:auto}
\eea
The  non-vanishing cross-correlations are
\bea
\,_{01}C^E_\ell(z_1,z_2)
&=&\frac{(\ell-1)(\ell+2)}{16\pi}\frac{(\ell+2)!}{(\ell-2)!}\int_0^{r_1}dr\frac{r_1-r}{rr_1}\int_0^{r_2}\frac{dr'}{r'}
\int k^2dk P_T(k,z,z')
\nonumber\\
&&\times
\frac{j_\ell(kr)}{(kr)^2}
\left[ \frac{j_{\ell+1}(kr')}{kr'}-(\ell+1)\frac{j_\ell(kr')}{(kr')^2}\right]
+\left( z_1 \leftrightarrow z_2 \right)
\nonumber\\
\,_{02}C^E_\ell(z_1,z_2)
&=&\frac{1}{16\pi}\frac{(\ell+2)!}{(\ell-2)!}\int_0^{r_1}dr\frac{r_1-r}{rr_1}
\int k^2dk P_T(k,z,z_2)
\nonumber\\
&&\times\frac{j_\ell(kr)}{(kr)^2}
\left[\frac{\ell(\ell -1)}{2} \frac{j_{\ell }(kr_2)}{(kr_2)^2}
-j_{\ell }(kr_2)+\frac{j_{\ell -1}(kr_2)}{kr_2}\right]
+\left( z_1 \leftrightarrow z_2 \right)
\nonumber\\
\,_{12}C^E_\ell(z_1,z_2)
&=&\frac{(\ell-1)(\ell+2)}{8\pi}\int_0^{r_1}\frac{dr}{r}
\int k^2dk P_T(k,z,z_2)
\left[ \frac{j_{\ell+1}(kr)}{kr}-(\ell+1)\frac{j_\ell(kr)}{(kr)^2}\right]
\nonumber\\
&&\times\left[\frac{\ell(\ell -1)}{2} \frac{j_{\ell }(kr_2)}{(kr_2)^2}
-j_{\ell }(kr_2)+\frac{j_{\ell -1}(kr_2)}{kr_2}\right]
+\left( z_1 \leftrightarrow z_2 \right)
\nonumber\\
\,_{12}C^B_\ell(z_1,z_2)
&=&-\frac{1}{8\pi}\int_0^{r_1}\frac{dr}{r}
\int k^2dk P_T(k,z,z_2)
\left[ (\ell^2-\ell-3)\frac{j_\ell (kr)}{kr}+ j_{\ell-1}(kr)+j_{\ell+1}(kr) \right]
\nonumber\\
&&\times\left[(\ell +2)\frac{ j_{\ell}(kr_2)}{kr_2}-j_{\ell +1}(kr_2)\right]
+\left( z_1 \leftrightarrow z_2 \right)\,.
\label{eq:cross}
\eea
Here $P_T(k,z_1,z_2)$ is the unequal redshift tensor power spectrum. It is obtained from the inital power spectrum $P_T(k)$ and the tensor transfer function $T_T(k,z)$ by $P_T(k,z_1,z_2)=P_T(k)T_T(k,z_1)T_T(k,z_2)$.
The relations between the comoving distance and redshift are obvious, $r_1=r(z_1)$, $r_2=r(z_2)$, $z=z(r)$ and $z'=z(r')$.
The $E$- and $B$-mode power spectra~\eqref{eq:power_spectrum}, determined through Eqs.~\eqref{eq:auto} and \eqref{eq:cross}), agree with the results derived in Refs.~\cite{Schmidt:2012ne,Schmidt:2012nw}, see Appendix~\ref{as:comp} for a detailed comparison.

In  Section~\ref{s:results}, we will present numerical results of our helicity decomposition for a tensor power spectrum from inflation.

\section{Convergence}
\label{s:convergence}
As already stressed before, our results show that in general $\kappa$ and $\ga_E$ are different. In this section, we explicitly evaluate the angular power spectrum of the convergence in order to quantify this difference for  scalar and tensor perturbations. To this aim, we start from the Jacobi map given in Eq.~\eqref{eq:cornerstone} and evaluate the convergence according to Eq.~\eqref{eq:39}. We find
\bea
\kappa(z,{\bf n})&=& \Psi
-\left(1-\frac{1}{\mathcal{H}r}\right)\delta z_{GI}
+\frac{\delta w_{GI}}{r}
+\frac{1}{2}\nabla_a\delta\theta^a_{GI}
-\frac{1}{2}H^{(t)}_{ab}\hat{\ga}^{ab}\,,
\label{eq:kappa-full}\\
\kappa^{(T)}(z,{\bf n})&=&
-\left(1-\frac{1}{\mathcal{H}r}\right)\delta w'_{GI}
+\frac{\delta w_{GI}}{r}
\nonumber\\
&&-\frac{1}{2}\int_0^{r_s}\frac{dr}{r^2}\left( 
\Delta_2\delta w_{GI}-2\,r\,\hat\ga^{ab}\nabla_aH^{(t)}_{rb} \right)
-\frac{1}{2}H^{(t)}_{ab}\hat{\ga}^{ab}\,.
\label{eq:kappa_tensor_1}
\eea
While the first line is the full $\kappa$, the second relation is valid only for tensor perturbations.

For comparison with the literature we also 
evaluate \eqref{e:dgaxt}
for scalar perturbations at fixed time, $t$, in longitudinal gauge. After some integration by part we find
\bea\label{e:kappaSt}
\kappa^{(S)}(t,{\bf n})&=& \psi +\int_0^{{r_z}}\frac{dr}{{r_z}}\left[\frac{{r_z}-r}{2\,r}\Delta_2(\Phi+\Psi)
 -(\Phi+\Psi)\right] \,,
 \eea
 where $r_z$ is the comoving distance out to redshift $z(t)$.
This result agrees with Ref.~\cite{Bernardeau:2009bm}, but note that it is not gauge invariant. Nevertheless, in addition to the traditional integral of $\Delta_2(\Phi+\Psi)$ which comes from $\nabla_a\delta\theta^a_{GI}$ and is also present in the scalar shear, there are additional relativistic terms.
The observable $\ka(z,\bn)$ contains additional corrections of similar amplitude as the relativistic terms.
For scalar perturbations Eq.~\eqref{eq:kappa-full} becomes
\bea\label{eq:kappa_scalar1}
\kappa^{(S)}(z,{\bf n})&=&
\Psi
-\left(1-\frac{1}{\mathcal{H}r}\right)\delta z_{GI}
+\frac{\delta w_{GI}}{r}
+\frac{1}{2} \nabla_a\delta\theta^a_{GI}\,,
\eea
where for scalar perturbations we have
\begin{align}
\delta w_{GI}=&-\int_0^r dr'\left( \Phi+\Psi \right)\,,
\nonumber\\
\nabla_a\delta\theta^a_{GI}=&\int_0^r dr'\frac{r-r'}{rr'}\Delta_2\left( \Phi+\Psi \right)\,,
\nonumber\\
\delta z_{GI}=&-\Phi
-\int_0^r dr'\left( \Phi'+\Psi' \right) 
-\pa_r\int_{0}^{t_s}dt\,\frac{a(t')}{a(t)}\,\Phi(t',r)\,,
\label{eq:deltaz}
\end{align}
so that finally
\bea\label{eq:kappa_scalar}
\kappa^{(S)}(z,{\bf n})&=&
\Psi+\Phi +\int_0^{{r_z}}\frac{dr}{{r_z}}\left[\frac{{r_z}-r}{2\,r}\Delta_2(\Phi+\Psi)
 -(\Phi+\Psi)\right] \nonumber \\  && +\left(1-\frac{1}{\mathcal{H}r}\right)\left[\int_0^r dr'\left( \Phi'+\Psi' \right) 
+\pa_r\int_{0}^{r_s}dr'\,\frac{a(t')}{a(t_s)}\,\Phi(t',r')\right] -\frac{1}{\mathcal{H}r}\Phi \,. \qquad \label{e:kappa-scal}
\eea
In $\La$CDM cosmology the term $\pa_r\int_{0}^{r_s}dr'\,\frac{a(t')}{a(t_s)}\,\Phi(t',r')$ can be replaced by the radial velocity, $V_{r}$. For this one has to assume that  sources and observer move on geodesics which is not necessarily true for the velocity field of the cosmic fluid which obeys the Euler equation. However, for pressureless matter in $\La$CDM the fluid velocity is geodesic. Comparing \eqref{e:kappa-scal} with \eqref{e:kappaSt}, we notice that several additional terms of the order of the gravitational potentials $\Psi$ appear as well as one term proportional to the velocity, $\pa_r\Psi$ which is parametrically a factor $k/\HH$ larger than potential terms.
Our result actually agrees up to the sign with the perturbation of the area distance which can be found e.g. in Refs.~\cite{Bonvin:2005ps,Marozzi:2014kua,durrer_2020}.
The expression for its power spectrum, the $C_\ell^{(s)\,\ka}(z_1,z_2)$ is somewhat cumbersome and not very illuminating. It is given explicitly in the Appendix, Eq.~\eqref{eq:scalar_conergence} and plotted in the figures of Section~\ref{s:results}.

Let us now concentrate on tensor perturbations.
We recall that for tensor perturbations
\be
\delta w_{GI}=\int_0^{r_s}dr\,H^{(t)}_{rr}\,.
\ee
At this point, we make use of the fact that $H^{(t)}_{ij}$ is transverse and traceless, so that we can write $r\nabla^a H^{(t)}_{ra}=-\left(\pa_r H^{(t)}_{rr}+\frac{3}{r}H^{(t)}_{rr}\right)$ and $\hat{\ga}^{ab}H^{(t)}_{ab}=-H^{(t)}_{rr}$. This leads to the following expressions for the convergence from tensor perturbations
\bea
\kappa(z,{\bf n})
&=&
-\left(1-\frac{1}{\mathcal{H}r}\right)
\int_0^{r_z}dr\,\pa_t H^{(t)}_{rr}
+\frac{1}{r_z}\int_0^{r_z}dr\,H^{(t)}_{rr}
\nonumber\\
&&-\frac{1}{2}\int_0^{r_z}dr\left( 
\frac{r_z-r}{r_zr}\Delta_2\,H^{(t)}_{rr}
+2\,\pa_r H^{(t)}_{rr}
+6\,\frac{H^{(t)}_{rr}}{r}\right)
+\frac{1}{2}H^{(t)}_{rr}
\nonumber\\
&=&
\int_0^{r_z}dr\,
\left[\frac{H^{(t)}_{rr}}{r_z}
-3\frac{H^{(t)}_{rr}}{r}
-\left(2-\frac{1}{\mathcal{H}r_z}\right)\pa_r H^{(t)}_{rr}
-\frac{1}{2} 
\frac{r_z-r}{r_zr}\Delta_2\,H^{(t)}_{rr}\right]
\nonumber\\
&&+\left(\frac{3}{2}-\frac{1}{\mathcal{H}r_z}\right)H^{(t)}_{rr}\,.
\label{eq:kappa_tensor_2}
\eea

Eq.~\eqref{eq:kappa_tensor_2} has been further manipulated in order to write all the integrated terms without time derivative. To this aim, we made use of the fact that all the integrals in Eq.~\eqref{eq:kappa_tensor_2} are done along the past light-cone, where  $\frac{d}{dr}=-\pa_t+\pa_r$. This result agrees (up to a sign) with the perturbation of the area distance from tensor perturbations, see~\cite{DiDio:2012bu}.
In Appendix~\ref{a:conv} we calculate the convergence power spectrum from \eqref{eq:kappa_tensor_2}. Here we just report the final result.

Introducing the kernels
\bea
\Ical_\ell(k,r,r')&=&\left(2-\frac{1}{\mathcal{H}r}\right)
\frac{j_{\ell +1}(kr')}{kr'}
\nonumber\\
&&+\left(
\frac{2-\ell(\ell+1)}{2}\frac{r'}{r}
-\frac{4(\ell -2)-\ell(\ell+1)}{2}
+\frac{\ell -2}{\mathcal{H}r}
-3
\right)\frac{j_{\ell }(kr')}{(kr')^2}
\nonumber\\
\Lcal_\ell(k,r)&=&\left(\frac{3}{2}-\frac{1}{\mathcal{H}r}\right)
\frac{j_{\ell }(kr)}{(kr)^2}\,,
\eea
the convergence  power spectrum from tensor perturbations is given by
\bea
C^\kappa_\ell(z_1,z_2)
&=&\frac{1}{8\pi}\frac{(\ell+2)!}{(\ell-2)!}
\int k^2 dk\left\{
\int^{r_1}_0\frac{dr'}{r'}
\int^{r_2}_0\frac{dr''}{r''}P_T(k,z',z'')
\Ical_\ell(k,r_1,r')\Ical_\ell(k,r_2,r'')\right.
\nonumber\\
&&+\left. P_T(k,z_1,z_2)\Lcal_\ell(k,r_1)\Lcal_\ell(k,r_2)
+\int_0^{r_1}\frac{dr'}{r'}P_T(k,z',z_2)\Ical_\ell(k,r_1,r')\Lcal_\ell(k,r_2)\right.
\nonumber\\
&&\left.+\int_0^{r_2}\frac{dr'}{r'}P_T(k,z_1,z')\Ical_\ell(k,r_2,r')\Lcal_\ell(k,r_1)\right\}\,.
\eea

The corresponding kernels $\Ical^E_\ell$ and $\Lcal^E_\ell$ for the shear $E$-modes are
\bea
\Ical^E_\ell(k,r,r')&=&
\left[\frac{\ell(\ell+1)}{2}\frac{r-r'}{r}
-(\ell -2)
-3\right]\frac{j_{\ell }(kr')}{(kr')^2}
+\frac{j_{\ell +1}(kr')}{kr'}
\nonumber\\
\Lcal^E_\ell(k,r)&=&\frac{1}{2}
\frac{j_{\ell }(kr)}{(kr)^2}\,,
\eea
which are clearly different. They are the kernels one obtains when considering only the term $\Delta_2\delta w_{GI}$ in \eqref{eq:kappa-full}.

While the difference between $\ka$ and $E$-mode shear power spectra for scalar modes are terms which are parametrically smaller, by at least one factor $(k/\HH)$ than the equal term which dominates on small scales, $k\gg \HH$, for tensor perturbations
the terms are all of the same order and we can at best say that $\ka$ and $E$-shear power spectra are of the same order of magnitude.

\section{Numerical results}
\label{s:results}

In this section we present the numerical evaluation of the convergence and the shear ($E$- and $B$- modes). We consider the Planck best fit cosmology~\cite{Planck:2018vyg}, as well as some simple analytical approximations.

\begin{figure}[!ht]
 \begin{center}
  \includegraphics[width=0.45\textwidth]{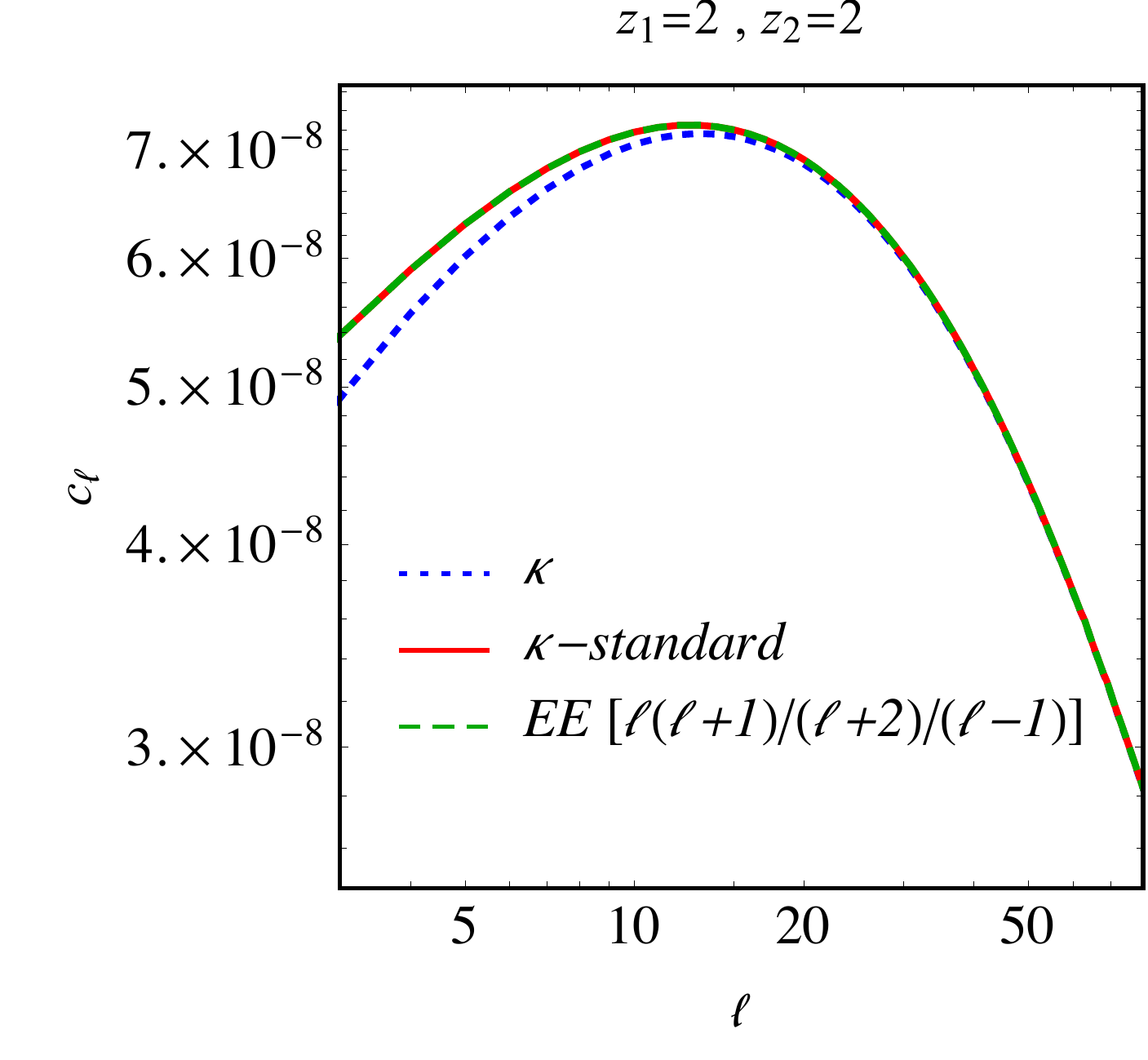}
  \hspace{1cm}
    \includegraphics[width=0.45\textwidth]{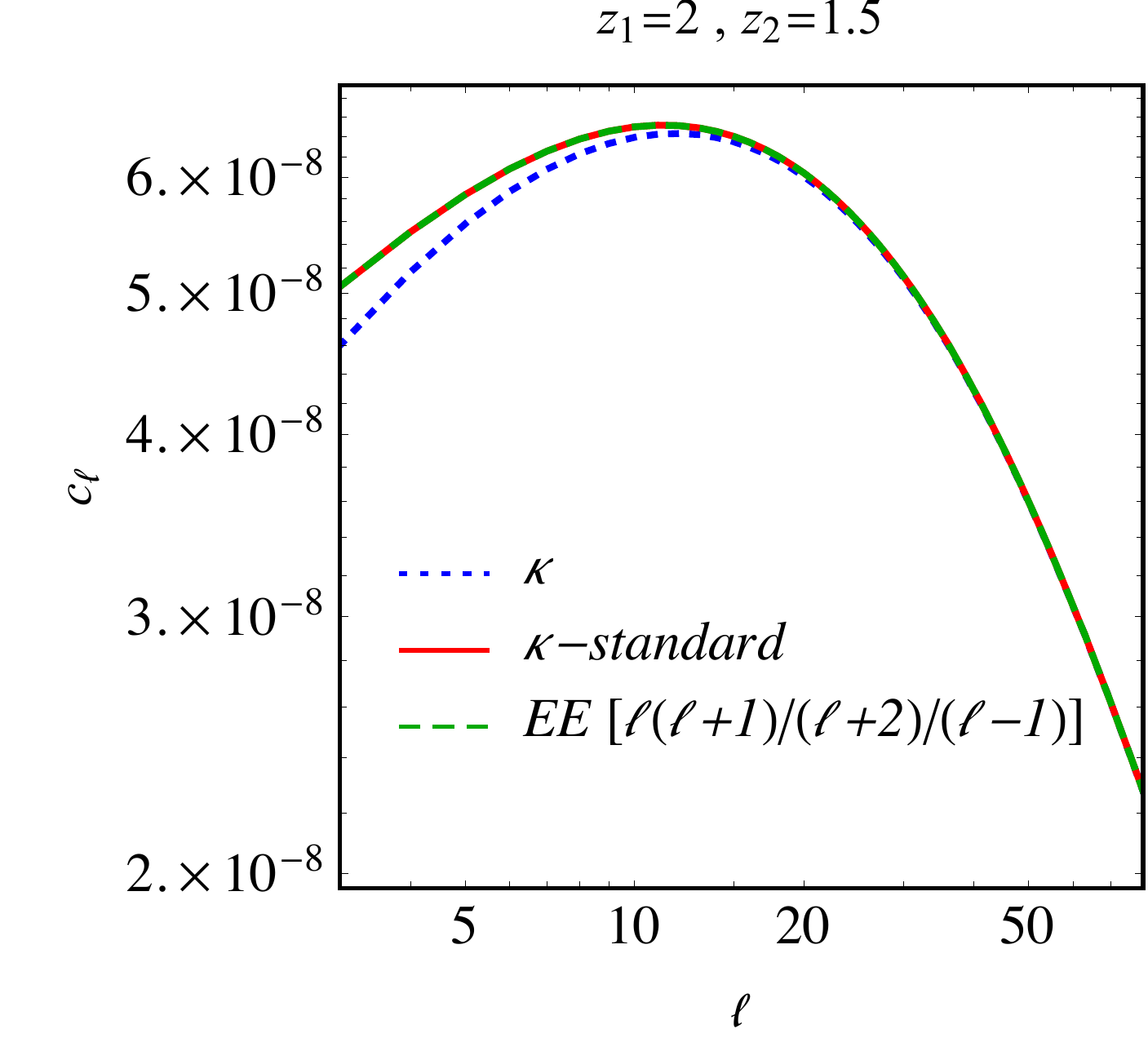}
\caption{We plot the convergence and shear $E$ modes sourced by scalar perturbations. On the left panel we consider sources at equal redshift ($z=2$), while on the right we cross-correlate sources at different redshifts ($z_1=2$ and $z_2=1.5$). We notice a difference at the largest scales, induced by the subleading lensing terms, which are not usually included in the standard lensing formalism.
}
\label{fig:comparison1}
 \end{center}
\end{figure}

In Fig.~\ref{fig:comparison1} we  plot the scalar $\kappa$ and $\ga_E$ power spectra where the latter is rescaled by a factor $\left[\ell(\ell+1)\right]^2/\left[(\ell+2)!/(\ell-2)!\right]$, so that it should agree with the $\ka$ power spectrum according to standard lensing calculations, see~\cite{Stebbins:1996wx,Hirata:2003ka,Schmidt:2012nw}. While this is an excellent approximation for $\ell\gtrsim 20$, it fails by up to 20\% at very low $\ell$'s.  It is an interesting open question whether this difference can be measured with future weak lensing and galaxy number count surveys. 

We now concentrate on tensor perturbations. The linear gravitational waves Fourier power spectrum is determined by
\be
P_T \left( k , t_1, t_2 \right) = T_h \left( k , t_1 \right) T_h \left( k , t_2 \right) P_{T0} \left( k \right)\,,
\ee
with
\be
P_{T0} \left( k \right) = 2 \pi^2 \frac{r A_s}{k^3} \left( \frac{ k}{k_0} \right)^{n_T}\,,
\ee
where $r$ denotes the scalar-to-tensor ratio at $k_0=0.002 \ {\rm Mpc}^{-1}$ and we chose $n_T= -r/8$ given by the consistency relation.
We have to determine the tensor transfer function which satisfies the equation
\be\label{e:tens'}
 T''_h + 2\mathcal{H}\, T'_h + k^2 T_h=0 \, ,
\ee
with the initial conditions $T_h \left( 0 \right) =1 $ and $T'_h \left( 0 \right)=0$.
In a matter-dominated universe the growing mode solution of \eqref{e:tens'} is given by
\be
T^{\rm M}_h = 3\,\frac{ j_1 \left( k t \right)}{k t} \, .
\label{eq:transfer_matter}
\ee
This is the approximation adopted in Ref.~\cite{Schmidt:2012nw}. However, this approximation neglects the phase of radiation domination which changes the initial conditions for the matter-dominated epoch. A more accurate approximation consists in considering first an era of radiation domination and  matching the initial conditions for the subsequent matter-dominated epoch. Following Refs.~\cite{Watanabe:2006qe,Saikawa:2018rcs,Kite:2021yoe} we obtain 
\bea
\label{eq:transfer_matterradiation}
T^{RM}_h= \left\{ \begin{array}{ccc} 
j_0 \left( k t \right) & \text{for} & t \le t_* \\
\frac{3}{k t} \left[ 
C j_1 \left( k t \right) 
- D y_1 \left( k t \right)\right] & \text{for} & t \ge t_*
\end{array} \right.
\eea
with
\bea
C&=&\frac{1}{2} - \frac{\cos \left( 2 k t_* \right)}{6} + \frac{\sin \left( 2 k t_* \right)}{3 k t_*} \, , \\
D&=& -\frac{1}{3 k t_*} + \frac{k t_*}{3} + \frac{\cos \left( 2 k t_* \right)}{3 k t_*} + \frac{\sin \left( 2 k t_* \right)}{6} \, , \\
t_* &=& \frac{4 \sqrt{\Omega_{\rm rel}}}{H_0 \Omega_M} \simeq 540 {\rm Mpc} \, .
\eea
Here $t_*$ is the conformal time at equal matter and radiation.

In addition to these two approximations we also consider  a universe without dark energy, $\Omega_\Lambda = 0$, where 
\be
\mathcal{H}^{\rm CDM} = \frac{1}{t} + \frac{1}{t+t_*}
\ee
and the complete solution in $\Lambda$CDM, where the Hubble parameter is computed with CLASS~\cite{Blas:2011rf}, with Planck best-fit cosmology~\cite{Planck:2018vyg}.

\begin{figure}[t]
 \begin{center}
  \includegraphics[width=0.45\textwidth]{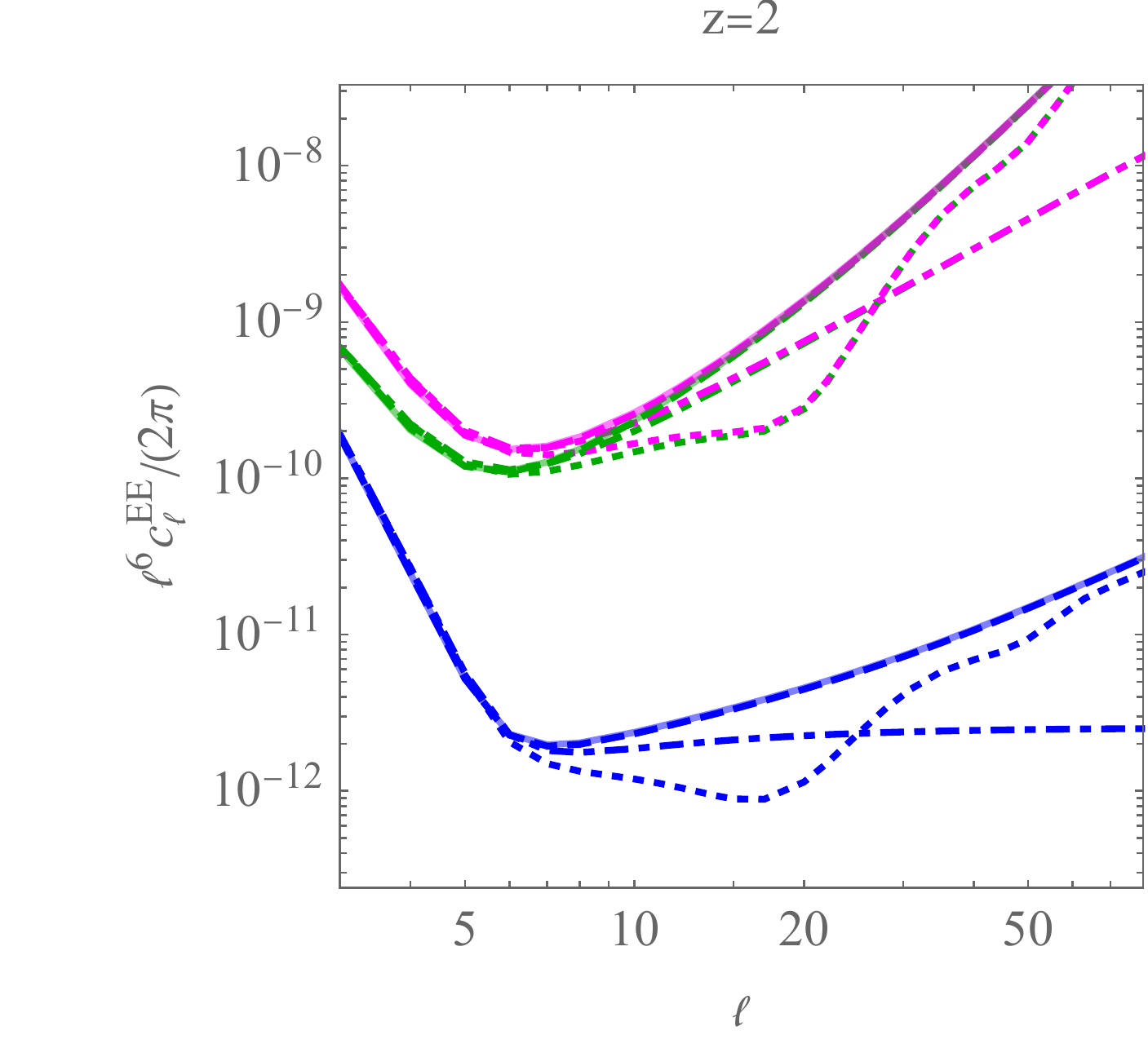}
  \hspace{1cm}
    \includegraphics[width=0.45\textwidth]{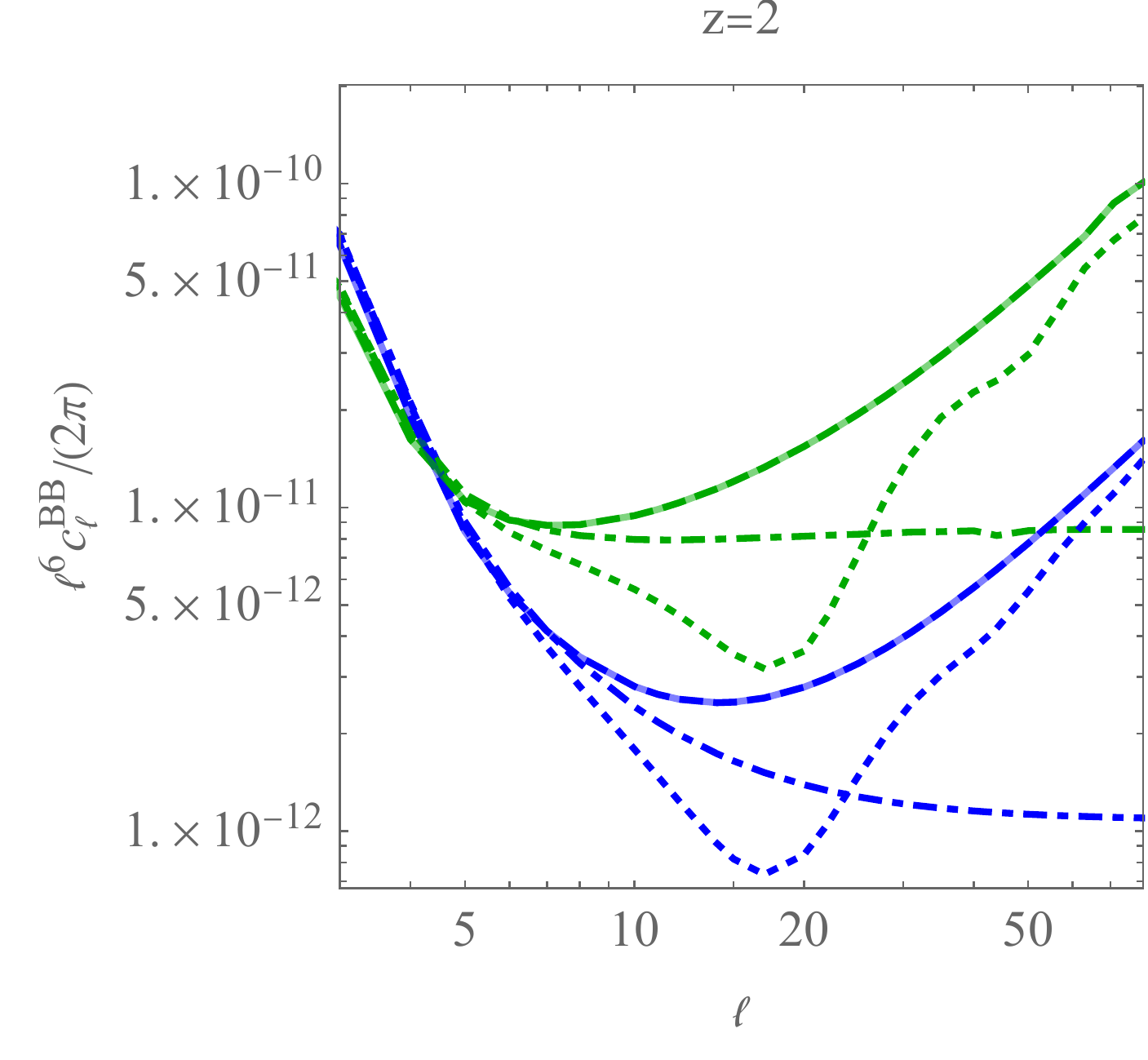}
\caption{We plot the shear $E$ (left) and $B$ (right) modes for tensor perturbations. We indicate the different contributions to Eqs.~(\ref{eq:SJ}) with different colors: blue for the terms involving $\hat Q_1$, green for $\hat Q_2$ and magenta for $\hat Q_3$. The different styles refer to the different approximations: dot-dashed for matter-dominated universe, dotted for radiation and matter, dashed for CDM and solid for $\Lambda$CDM.
To compare with Ref.~\cite{Schmidt:2012nw} we use $r=0.2$. 
}
\label{fig:comparison}
 \end{center}
\end{figure}

\begin{figure}[!ht]
 \begin{center}
  \includegraphics[width=0.5\textwidth]{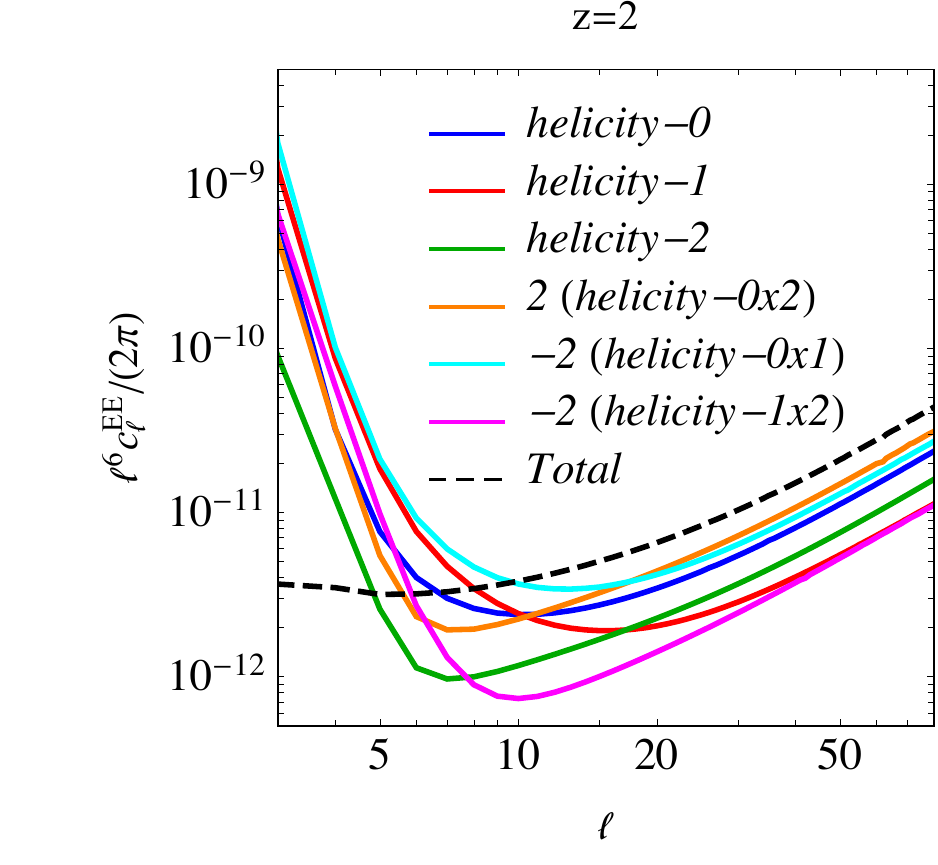}~\includegraphics[width=0.5\textwidth]{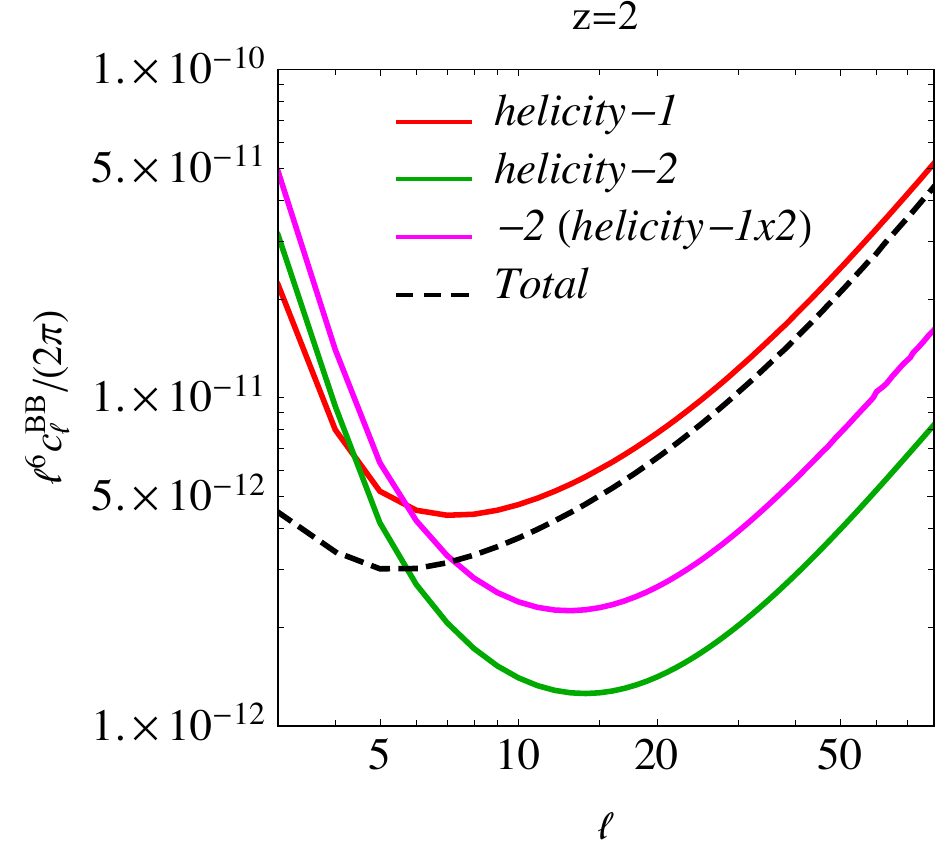}
  \caption{We plot the shear $E$ (left) and $B$ (right) modes. With different colors we indicate the different contributions, as defined in Eqs.~(\ref{eq:auto}, \ref{eq:cross}). We use the scalar-to-tensor ration $r=0.1$.
 }
\label{fig:auto}
 \end{center}
\end{figure}

\begin{figure}[!ht]
 \begin{center}
  \includegraphics[width=0.48\textwidth]{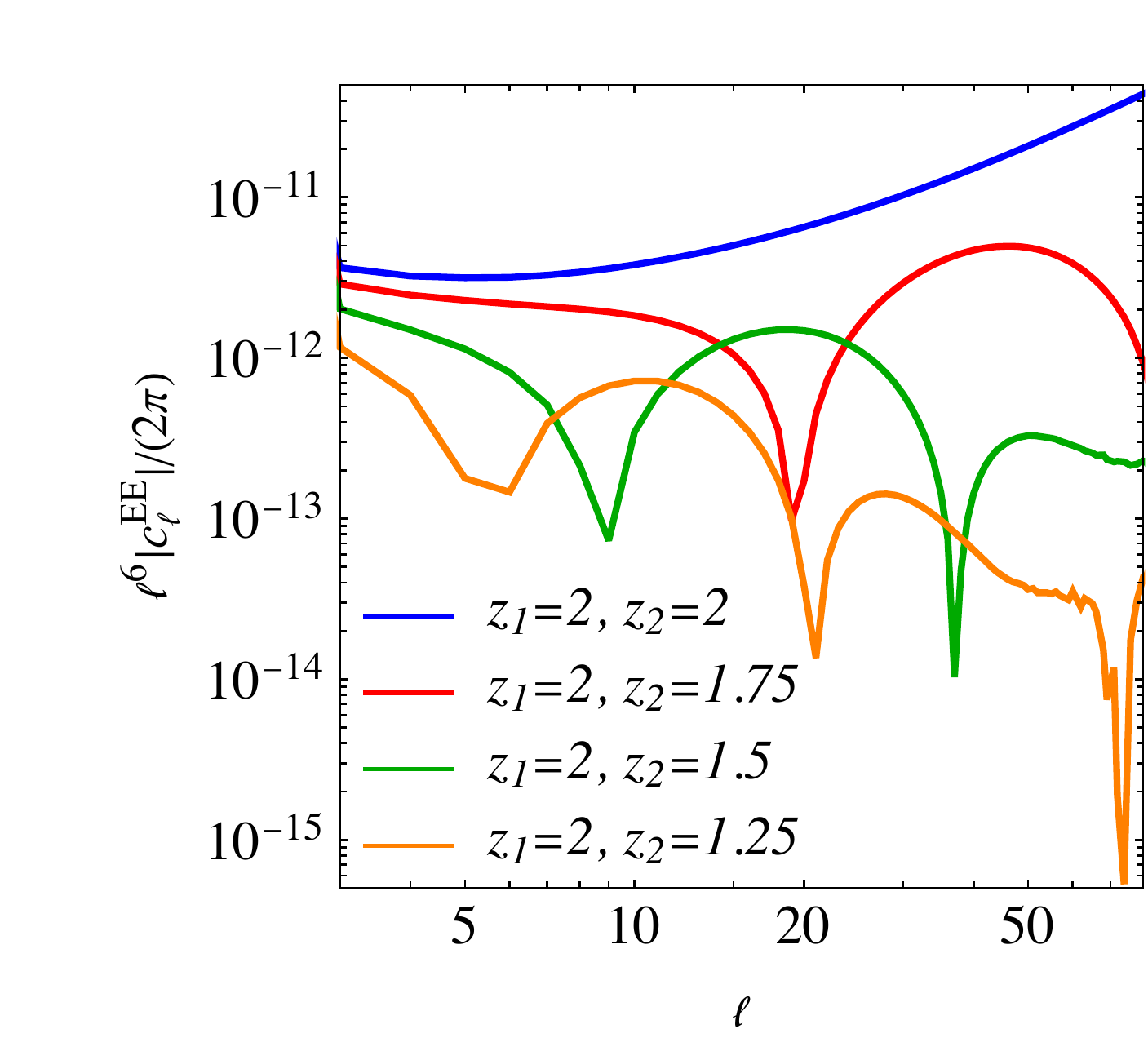}~~\includegraphics[width=0.48\textwidth]{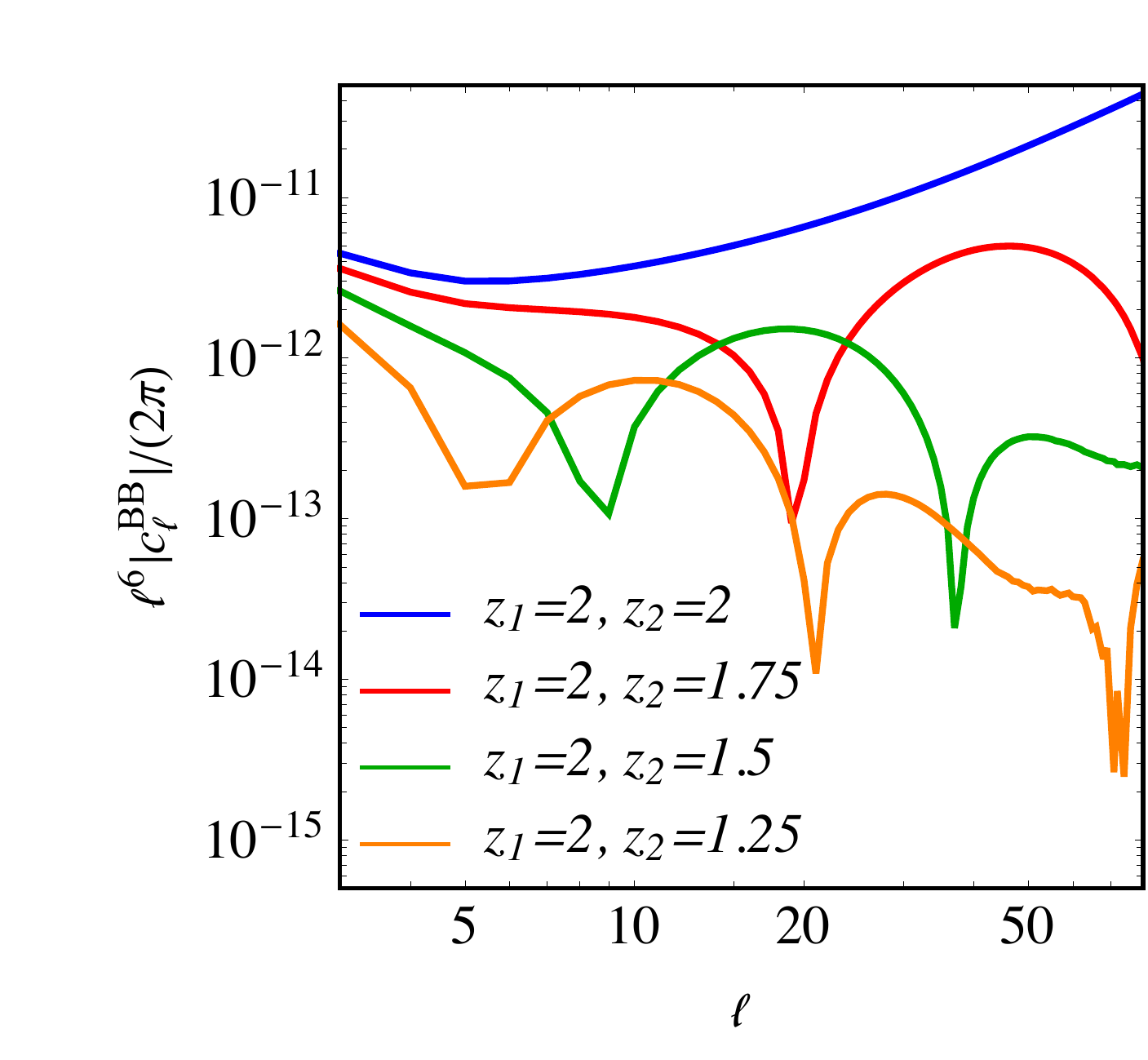}\\
    \includegraphics[width=0.48\textwidth]{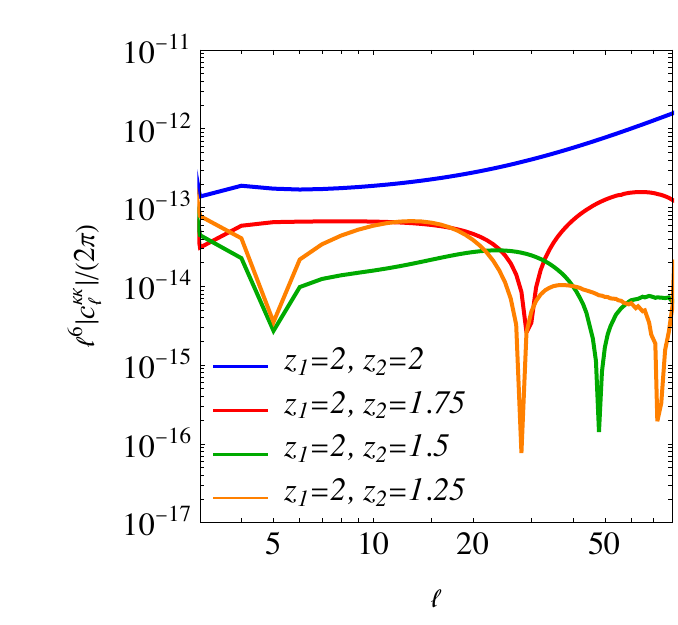}
    ~~\includegraphics[width=0.48\textwidth]{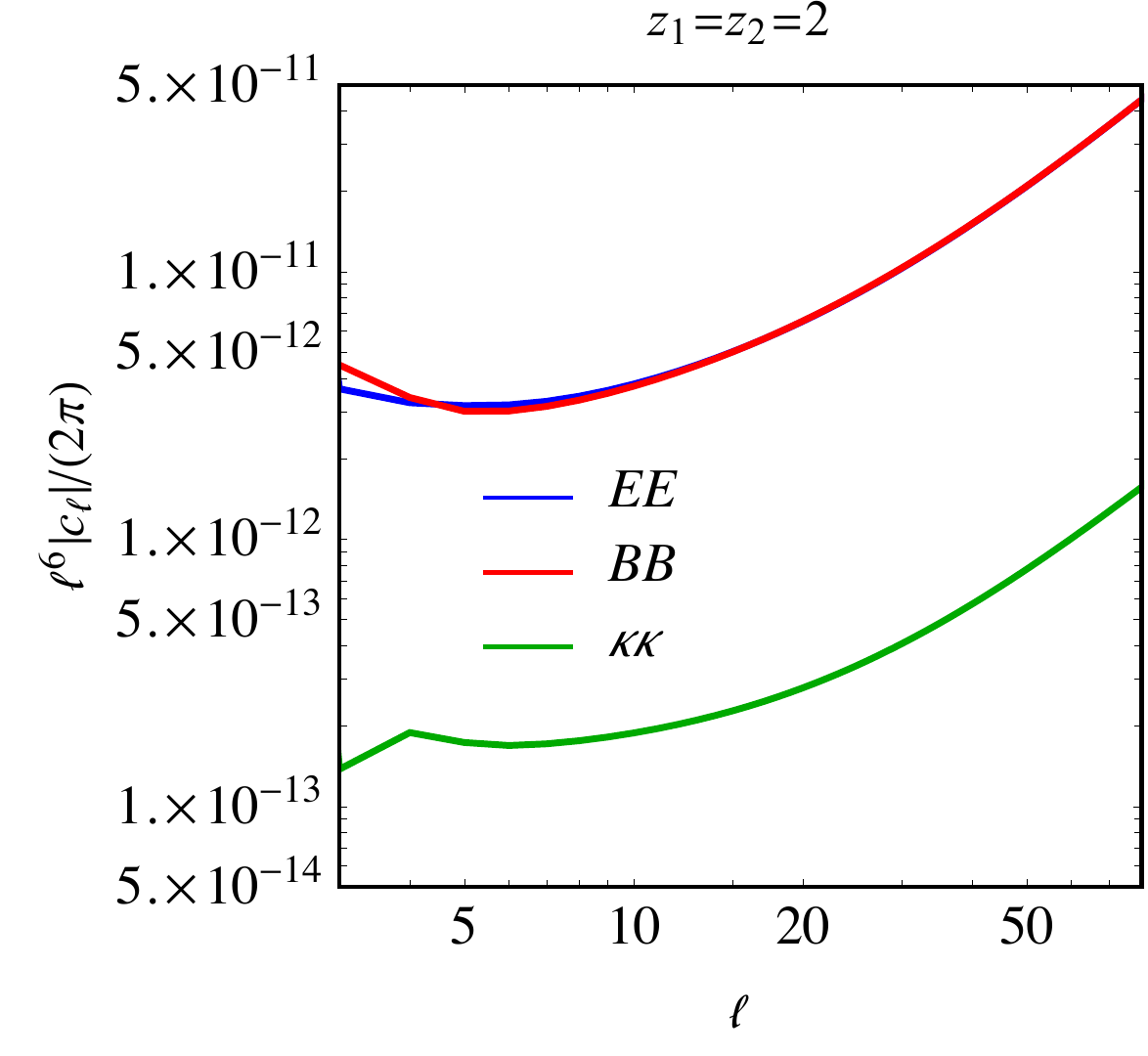}
    \caption{We plot the shear $E$ (top-left) and $B$ (top-right) modes and $\kappa$ (bottom-left). Different colors refer to  absolute value of different redshift cross-correlation. In the bottom-right plot we compare the power spectra for $z_1=z_2=2$. The $E$- and $B$-spectra agree for $\ell>6$ and the convergence spectrum is about a factor 50 lower than the shear. We use the scalar-to-tensor ration $r=0.1$.
\label{fig:cross}}
 \end{center}
\end{figure}

\begin{figure}[!ht]
 \begin{center}
  \includegraphics[width=0.48\textwidth]{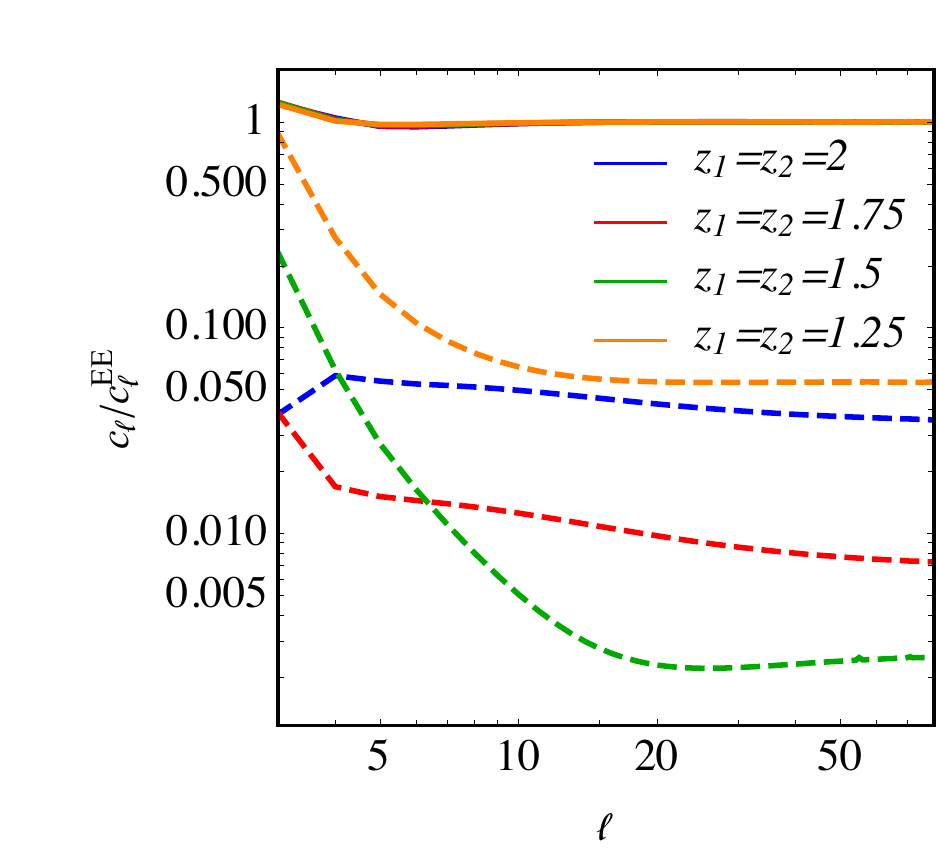}
  ~~
    \includegraphics[width=0.48\textwidth]{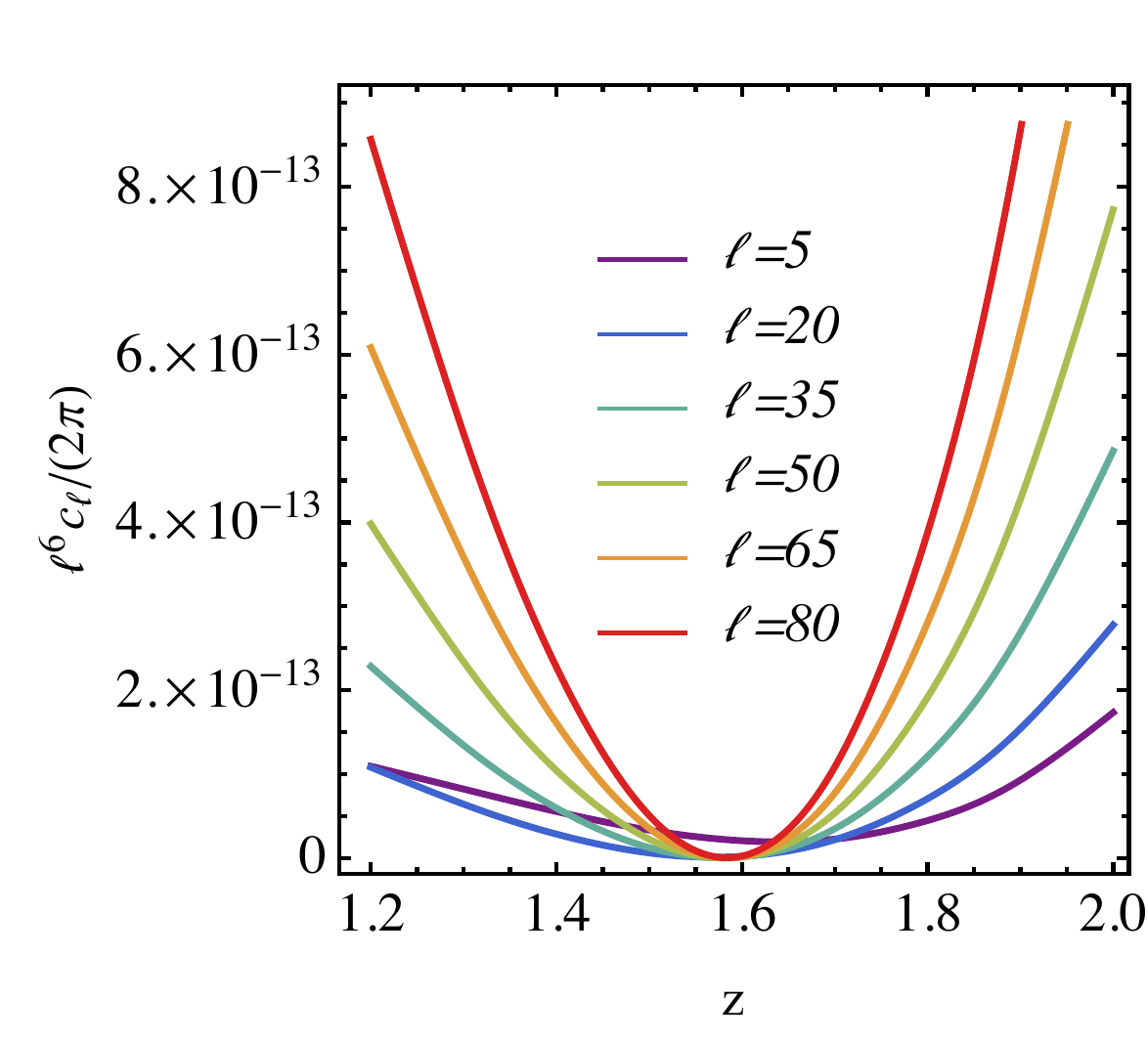}
  \caption{ On the left panel we plot the ratio between shear $E$ and $B$ modes (solid lines) and between shear E modes and the convergence $\kappa$ (dashed lines) at different redshifts.
  On the right panel we show the redshift evolution of $\kappa$ for different multipoles. We remark that its behaviour is not monotonic with the redshift $z$. We remark that this in line with the maximum of the angular diameter distance $z_*\simeq 1.59$, as already pointed out in Ref.~\cite{DiDio:2012bu}}
\label{fig:ratio}
 \end{center}
\end{figure}

\begin{figure}[!ht]
 \begin{center}
    \includegraphics[width=1.05\textwidth]{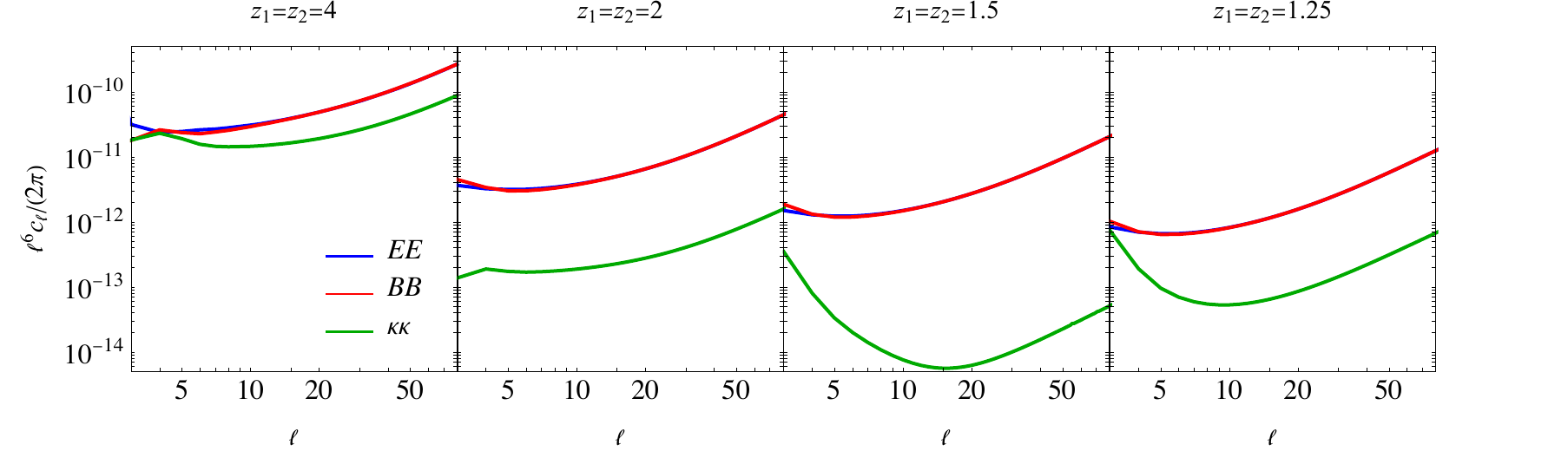}
  \caption{We plot the spectra for shear $E$ and $B$ modes, together with the convergence $\kappa$ at different redshifts.}
\label{fig:all_z}
 \end{center}
\end{figure}

In Fig.~\ref{fig:comparison} we compare the contributions from $\hat Q_1$, $\hat Q_2$ and $\hat Q_3$ to the shear $E$- and $B$- mode power spectra. These quantities are introduced in Appendix~\ref{as:comp}, see~\cite{Schmidt:2012ne,Schmidt:2012nw} for details. We plot these quantities for easy comparison with Ref.~\cite{Schmidt:2012nw}. All our plots start at $\ell=3$ since our expressions formally diverge at $\ell=2$. This is due to the fact that we have neglected the term at the observer which regularizes this unphysical divergence, see~\cite{Schmidt:2012nw} for a discussion. We see that except at the largest scales, $\ell\lesssim 10$, the two approximations~\eqref{eq:transfer_matter}-(dot-dashed lines used in Ref.~\cite{Schmidt:2012nw}) and \eqref{eq:transfer_matterradiation} (dotted lines) fail to describe correctly scales that enter the horizon during the radiation era.
The simplest approximation, Eq.~\eqref{eq:transfer_matter}, agrees well with the results presented in Ref.~\cite{Schmidt:2012nw}.
The more accurate approximation provided by Eq.~\eqref{eq:transfer_matterradiation} is quite good at smaller scales, failing therefore only on the scales comparable with the transition between the two regimes.
Instead the impact of $\Omega_\Lambda$ on the Hubble parameter does not lead to any sizable effect and the pure CDM spectra cannot be distinguished from the $\La$CDM ones. Of course at $z=2$ the universe is matter dominated and $\La$ is subdominant. But since lensing is an integrated quantity, also low redshifts  contribute to it. Nevertheless, these contributions cannot alter the signal appreciably. We have also computed the lensing signal at low redshift, $z=0.3$ where $\La$ is still dominant. But also there the difference between the CDM transfer function and the correct $\La$CDM one changes the (much smaller) result only by about 20\%. Hence tensor-shear is not a useful observable to measure $\La$.

In fig.~\ref{fig:auto} we show the absolute value of the different helicity  contributions to the $E$- and $B$- modes of the shear as defined in Eqs.~(\ref{eq:auto},~\ref{eq:cross}) . We see that all the terms contribute with similar amplitudes and shapes to the total power spectrum. Note also that on large scales, the negative cross correlations of the different helicities lead to significant cancellation, while on smaller scales, $\ell>10$ their contribution is less relevant. Both the $E$-mode and $B$-mode power spectra have the same amplitudes for $\ell>6$. We have checked numerically that this equality persists also for higher $\ell$. In Appendix~\ref{a:evolution} we show that the evolution equation for the $E$- and $B$-modes of the shear is governed by the $E$ and $B$ parts of the Weyl tensor. We also show that for statistically isotropic, free gravitational waves inside the horizon, $k>\HH$, the  $E$ and $B$ parts of the Weyl tensor have the same power spectrum.

In Ref.~\cite{Schmidt:2012nw} it has been shown that this tensor auto-correlation spectrum is much smaller than intrinsic alignment and will therefore be very hard to measure. For this reason we also show the cross-correlations for different redshifts which are not plagued by intrinsic alignment.

In fig.~\ref{fig:cross} we consider the correlations between different redshifts. The coefficients of the expansion in spherical harmonics of the $E$- and $B$- modes oscillate with a frequency determined by the comoving distance to the source. By correlating different redshifts we therefore multiply spherical Bessel functions with different phases which lead to oscillations in the cross power spectrum. Moreover, the conversion from the angular multipole $\ell$ to the physical separation between the two sources is redshift dependent. In particular, a given multipole $\ell$ corresponds to a larger physical distance when increasing the redshift difference $\Delta z = \left| z_1 - z_2 \right|$, leading to a suppression of the angular power spectrum. For the largest redshift difference, $z_2=1.25$ (orange lines) the rapid oscillations of the spherical Bessel functions are challenging our accuracy. The wiggles at $\ell>60$ are not physical but an accuracy problem.

In fig.~\ref{fig:cross}, lower panel, we also show the tensor spectrum of the convergence $\ka$. While the standard treatment of lensing predicts that this should agree with the shear $E$-mode spectrum (up to an $\ell$-factor which rapidly converges to $1$ for increasing $\ell$), we see a very different behavior. At equal redshift (blue line) the $\ka$ power spectrum is by more than an order of magnitude smaller than the shear $E$-mode power spectrum. This remains roughly true for the unequal redshift spectra even though the shapes, e.g., the number and positions of the zero-crossings, are similar. 

Contrary to that, the $E$- and $B$-mode shear spectra are identical for $\ell>6$ within numerical accuracy both for equal and unequal redshift spectra. This equality comes from the fact that they are generated respectively by the $E$- and $B$-part of the Weyl tensor of the gravitational waves. In Appendix~\ref{a:evolution} we show this and we also show that the $E$- and $B$-parts of the Weyl tensor for free gravitational waves inside the horizon have identical spectra.

In Fig.~\ref{fig:ratio} we show the ratio between the shear E- and B-modes which is always very close to one and the ratio between the convergence and the shear E-mode which has a minimum of about $10^{-3}$ at $z=z_* \simeq 1.6$ where the angular diameter distance is maximal. At higher redshifts $\ka$ increases again and approaches the shear spectra for 
$z\gtrsim 4$, see Fig.~\ref{fig:all_z}.

Our results can be summarized as follows: For linear tensor perturbations, the $E$- and $B$-mode spectra of the shear are identical, the $\ka$-spectrum is more than an order of magnitude smaller for $z\leq 3$ and the rotation vanishes. This is in contrast to scalar perturbations where at sufficiently high $\ell$ the shear $E$-mode and the $\ka$ spectra agree while the shear $B$-mode and the rotation vanish.

\section{Conclusions}\label{s:con}
In this paper we have computed the Jacobi map to first order in cosmological perturbation theory. This is the correct gauge invariant and fully relativistic lensing map. We have shown that contrary to the often used 'amplification matrix' $\pa_a\theta_b$, the Jacobi matrix has in general four degrees of freedom and is symmetric at first order in perturbation theory also when vector and tensor perturbations are included. Therefore lensing rotation vanishes at first order also for vector and tensor perturbations. Furthermore, the simple relations between the convergence $\ka$ and the shear $E$-mode $\ga_E$ are not valid. While for scalar perturbations the difference is relatively small (up to 20\%) and only relevant on large scales, $\ell\lesssim 20$, it is very significant for tensor perturbations, for which the corresponding power spectra differ by more than an order of magnitude. More precisely, the tensor convergence power spectrum is on average a factor 10 to 50 times smaller then the tensor shear $E$-mode power spectrum. This factor depends on the redshift and is largest at $z_*\simeq 1.6$ which maximizes the angular diameter distance. Even though, at first, this is surprising, it is certainly not inconceivable. While for scalar perturbations there is a hierarchy between time derivatives, $\propto \HH$ and spatial derivatives $\propto k$, this is no longer the case for tensor perturbations which oscillate with frequency $k$. Therefore, the additional terms which appear in our relativistic treatment, which contain radial derivatives that can be converted into time derivatives on the lightcone, are expected to be of the same order of magnitude as the original terms, and therefore they can change the result very substantially.

This equivalence of spatial and temporal derivatives for free, subhorizon tensor perturbations also explains the fact that the $E$- and $B$-mode shear power spectra are identical well inside the horizon, for $z=2$ this is above $\ell\sim 6$.

Furthermore, the simple relation between the $\ga_B$ and rotation power spectrum is not valid as the latter vanishes at first order in perturbation theory while the former is equal to the $E$-mode spectrum for tensor perturbations.

While lensing shear spectra at equal redshift are plagued by intrinsic alignment and therefore difficult to measure, cross spectra at different redshift might be a way out of this systematic problem.

\section*{Acknowledgements}
The authors wish to thank Daniel Thomas for insightful discussions that motivated this work.
GF acknowledges support by Funda\c{c}\~{a}o para a Ci\^{e}ncia e a Tecnologia (FCT) under the program {\it ``Stimulus"} with the grant no. CEECIND/04399/2017/CP1387/CT0026 and through the research project with ref. number PTDC/FIS-AST/0054/2021. GF is also member of the Gruppo Nazionale per la Fisica Matematica (GNFM) of the Istituto Nazionale di Alta Matematica (INdAM).
RD thanks the University of Pisa for hospitality and financial support under the Visiting Fellow program. RD also thanks the Swiss National Science Foundation for financial support. GM is supported in part by INFN under the program TAsP ({\it Theoretical Astroparticle Physics}).

\appendix

\section{The perturbed angular metric in GLC gauge}\label{a:per-ga}
\subsection{Coordinate transformation from GLC gauge to standard linear theory}
\label{sec:A1}
In this appendix we derive the expression for the GLC $\de\ga_{ab}$ in terms of general metric perturbations as defined in Eqs.~\eqref{e:met1} to \eqref{e:met3}.  A  coordinate transformation between GLC metric $g^{GLC}_{\mu\nu}$ \eqref{4} and the a metric $g_{\alpha\beta}$ in Eq.~\eqref{e:met1} is given by
\be
g^{\mu\nu}_{GLC}=\frac{\pa x^\mu}{\pa y^\alpha}\frac{\pa x^\nu}{\pa y^\beta} g^{\alpha\beta}\,.
\label{eq:coord_tr}
\ee
Here $x^\mu=(\tau+\delta\tau,w+\delta w,\theta^a+\delta\theta^a)$, $y^\alpha=(t,x^i)$ and $g^{\mu\nu}_{GLC}$ and $g^{\alpha\beta}$ are respectively the linearized inverse of $g^{GLC}_{\mu\nu}=\bar g_{\mu\nu}+\delta g_{\mu\nu}$ and $g_{\alpha\beta}=\bar g_{\alpha\beta}+\delta g_{\alpha\beta}$. The the $\tau\tau$, $ww$ and $wa$ components of Eq.~\eqref{eq:coord_tr} then lead to the following equations for the GLC coordinates,
\bea
\delta\tau'&=&a\,\phi\nonumber\\
\frac{d\delta w}{d\lambda}&=&\phi-n^i n^j H_{ij}-n^i B_i\nonumber\\
\frac{d\delta\theta^a}{d\lambda}&=&\delta^{ij}\pa_i\theta^a\left( \pa_j\delta w-B_j-2\,n^k H_{kj} \right)\,,
\label{eq:expl_coord_tr}
\eea
where $'\equiv \pa_t$,  $\frac{d}{d\lambda}\equiv \pa_t-n^i\pa_i$ and 
we have defined $r=\sqrt{\delta_{ij}x^i x^j}$, $n^i\equiv x^i/r$ as well as $\theta^a\equiv\left( \arccos\left( \frac{x^2}{x^1} \right),\arctan\left( \frac{x^3}{r} \right) \right)$.
It is important to note that $\frac{d}{d\lambda}$ is the total derivative along the background light cone, i.e. in direction of the null-vector $(k^\mu) =(1,n^i)$, whereas a prime $'$ is the partial  derivative w.r.t.~conformal time at fixed space coordinates.
By applying the SVT decomposition in Eqs.~\eqref{e:met2} and \eqref{e:met3} and  defining the Bardeen potentials $\Phi$ and $\Psi$ as
\begin{equation}
\Phi=\phi-\left( B+E' \right)'-\mathcal{H}(B+E')\qquad\text{and}\qquad\Psi=\psi+\mathcal{H}(B+E')+\frac{1}{3}\Delta E\,,
\end{equation}
we obtain
\beq
\delta\tau'=a\,\Phi+\left[a\left( B+E' \right)\right]'\,,
\eeq
which is solved by
\beq
\delta\tau=\int_{0}^{t_s}dt\,a\,\Phi+a\left( B+E'\right)= \delta\tau_{GI}+a\left( B+E'\right)\,,
\label{eq:delta_tau}
\eeq
where we have introduced
\be
\delta\tau_{GI}\equiv\int_{0}^{t_s}dt\,a\,\Phi\,.
\ee
From now on, the subscript $s$ refers to a quantity evaluated at the source position. In this way, the second of Eqs.~\eqref{eq:expl_coord_tr} can be written as
\be
\frac{d\delta w}{d\lambda}
=\frac{d \left( n^i H^{(v)}_i+B+n^j E_{,j}+E'\right)}{d\lambda}+\Phi+\Psi-n^i \sigma^{(v)}_i-n^i n^j H^{(t)}_{ij}\,,
\ee
where $\sigma^{(v)}_i=B^{(v)}_i+{H^{(v)}_i}'$ is a gauge invariant vector perturbation. This leads to
\bea
\delta w&=&
-\int_0^{r_s}dr\left(\Phi+\Psi -n^i\sigma_i^{(v)}-n^i n^j H^{(t)}_{ij} \right)+n^iH^{(v)}_i+B+n^j E_{,j}+E'
\nonumber\\
&=&\delta w_{GI}+n^i\chi_i+B+E'\,,
\label{eq:delta_w}
\eea
where we have defined $\chi_j\equiv H^{(v)}_j +E_{,\,j}$ and
\be
\delta w_{GI}\equiv
-\int_0^{r_s}dr\left(\Phi+\Psi -n^i\sigma_i^{(v)}-n^i n^j H^{(t)}_{ij} \right)(t_o-r,r,\theta^a)\,.
\ee
Here $t_o$ denotes present time.
Finally, the last of Eqs.~\eqref{eq:expl_coord_tr} gives
\bea
\frac{d\delta\theta^a}{d\lambda}
&=&\delta^{ij}\pa_i\theta^a\left( \pa_j\delta w_{GI}
+\pa_j n^kH^{(v)}_k
-B^{(v)}_j
-{H^{(v)}_j}'
+\frac{d H^{(v)}_j}{d\lambda}
-2\,n^k H^{(t)}_{kj}
+\pa_j n^k E_{,k}
+\frac{d E_{,j}}{d\lambda}
\right)\nonumber\\
&=&\frac{d\delta\theta^a_{GI}}{d\lambda}
+\delta^{ij}\pa_i\theta^a\left(
\pa_j n^k\,\chi_k
+\frac{d \chi_j}{d\lambda} \right)\,,
 \label{eq:angles}
\eea
where we have introduced
\bea
\delta\theta^a_{GI}&\equiv&
\int_0^{\lambda_s}d\lambda\delta^{ij}\pa_i\theta^a\left( 
\delta w_{GI\,,j}-\sigma^{(v)}_j-2 n^k H^{(t)}_{kj} \right)
\nonumber\\
&=&-\int_0^{r_s}dr\delta^{ij}\pa_i\theta^a\left( 
\delta w_{GI\,,j}-\sigma^{(v)}_j-2 n^k H^{(t)}_{kj} \right)(t_o-r,r,\theta^a)\,,
\eea
and we have used the geometrical identity $n^i\pa_i\theta^a=\pa_r\theta^a=0$. In order to simplify the gauge dependent term in Eq.~\eqref{eq:angles}, let us note that
\be
\delta^{ij}\pa_i\theta^a\frac{d \chi_j}{d\lambda}
=\frac{d\left( \delta^{ij}\pa_i\theta^a \chi_j \right)}{d\lambda}
-\delta^{ij}\frac{d \pa_i\theta^a}{d\lambda}\chi_j
=\frac{d\left( \delta^{ij}\pa_i\theta^a \chi_j \right)}{d\lambda}
+\delta^{ij}n^k\pa_k\left( \pa_i\theta^a\right)\chi_j\,,
\ee
since $\pa_i\theta^a$ does not depend on $t$. Now we have that the combination
\beq
\delta^{kj}\left( \pa_i\theta^a\pa_j n^i+n^i\pa_{ij}\theta^a\right)=
\delta^{kj}\pa_j\left( n^i\pa_i\theta^a \right)=0\,,
\eeq
since $n^i\pa_i\theta^a=\pa_r\theta^a=0$. With this,  the deflection angle can finally be written as
\be
\delta\theta^a
=\delta\theta^a_{GI}+\delta^{ij}\pa_i\theta^a\,\chi_j\,.
\label{eq:delta_theta}
\ee

Eqs.~\eqref{eq:delta_tau}, \eqref{eq:delta_w} and \eqref{eq:delta_theta} above share the following property: the integrated terms are gauge invariant, as expected, whereas the gauge dependent part can be entirely taken out of the integral and can be written as a boundary term. This is a consequence of the fact that a gauge transformation can be written as a Lie derivative, acting as a differential operator rather than an integral. Let us also note that $\pa_i\theta^a$ is precisely the angular part of the Jacobian needed to transform from carthesian to polar coordinates. Since this transformation preserves the $SO(3)$ background symmetry and both $H^{(v)}_j$ and $E_{,\,j}$ are vectors under this symmetry group, we can immediately write
\be
\delta w= \delta w_{GI}+\chi_r+B+E'
\qquad\text{and}\qquad
\delta\theta^a=\delta\theta^a_{GI}+\chi^a\,.
\label{eq:polar_gauge_dependence}
\ee
Finally, let us compute the perturbation of the inverse $\gamma^{ab}=\bar\gamma^{ab} +\de\ga^{ab}$ to first order, needed for the expression of the Jacobi map. To this aim, we make use of the $ab$ component of Eq.~\eqref{eq:coord_tr}. The background term is given by
\beq
\bar{\gamma}^{ab}=a^{-2}\delta^{ij}\pa_i\theta^a\pa_j\theta^b\,.
\eeq
After some straightforward manipulations, also the linear term can be written as
\bea
\delta\gamma^{ab}
&=&a^{-2}\left(\pa_i \delta\theta^a_{GI} \pa_j \theta^b \delta^{ij}
+\pa_i \theta^a \pa_j \delta\theta^b_{GI} \delta^{ij}
-2\,\pa_i \theta^a \pa_j \theta^b \delta^{ik}\delta^{jl}\,H^{(t)}_{kl}\right)\nonumber\\
&&+\delta^{kl}\pa_k\bar\gamma^{ab} \chi_l
+2\,\bar\gamma^{ab}\left( \psi+\frac{1}{3}\Delta E \right)\nonumber\\
&=&\delta\gamma^{ab}_{GI} ~ 
+\delta^{kl}\pa_k\bar\gamma^{ab} \chi_l
-2\,\mathcal{H}\,\bar\gamma^{ab}\left( B+E' \right)
\label{eq:delta_gamma}
\eea
where we have defined 
\beq
\delta\gamma^{ab}_{GI}\equiv a^{-2}\left(\pa_i \delta\theta^a_{GI} \pa_j \theta^b \delta^{ij}
+\pa_i \theta^a \pa_j \delta\theta^b_{GI} \delta^{ij}
-2\,\pa_i \theta^a \pa_j \theta^b \delta^{ik}\delta^{jl}\,H^{(t)}_{kl}\right)
+2\,\bar\gamma^{ab}\Psi\,.
\eeq

We therefore find that also $\delta\gamma^{ab}$ is  not gauge invariant. This is in agreement with what has been found when  perturbation theory is applied directly in the light-cone coordinates~\cite{Fanizza:2020xtv,Mitsou:2020czr,Frob:2021ore}. This seems to suggest that not even the Jacobi map at linear order is gauge invariant. However, we show in the following part of this Appendix that this gauge dependence  just come from the fact that background coordinates are left unspecified. Gauge invariance will be indeed restored once the Jacobi map is expressed as a function of  observable quantities like the redshift on the observer's past light-cone and the incoming directions of photons.

\subsection{Observed redshift  and incoming photon directions on the past light-cone}
\label{sec:A2}
We want to express the perturbed Jacobi map in terms of the fiducial model background coordinates where the time coordinate is replaced by the observed redshift and the observed position of source is given by the observer direction of the incoming photons and their redshift. This provides observable background coordinates on the past lightcone. Let us first determine the perturbed redshift of the source. With the normalization $\left(k_\mu u^\mu\right)_o=-a_o=-1$ we  have
\bea \label{eq:z_pert}
1+z&=&-\left(k^\mu u_\mu\right)_s=a^{-1}-\delta\left(g^{\mu\nu}\pa_\mu\tau\pa_\nu w\right)
\nonumber\\
&=&a^{-1}\left[1
-\Phi
+n^{j}\sigma^{(v)}_j
+\delta w'_{GI}
-a^{-1}n^i\pa_i\delta\tau_{GI}
-\mathcal{H}\left( B+E' \right)\right]
\nonumber\\
&=&a^{-1}\left[1
+\delta z_{GI}-\mathcal{H}\left( B+E' \right)\right]\,,
\eea
where we have introduced 
\be
\delta z_{GI}\equiv
-\Phi
+n^{j}\sigma^{(v)}_j
+\delta w'_{GI}
-a^{-1}n^i\pa_i\delta\tau_{GI}\,.
\ee
Since $a^{-1} = 1+\bar z$, 
\be
\frac{\delta z}{1+\bar z}= \delta z_{GI}-\mathcal{H}\left( B+E' \right)\,.
\label{eq:redshift_gauge_dependence}
\ee
We are now in the position to fix the gauge dependence of $\delta\gamma^{ab}$ by expressing it in terms of the observed redshift and incoming directions on the past light-cone. Clearly, observable quantities expressed in terms of other observables cannot depend on the coordinate system. Indeed, Eqs.~\eqref{eq:polar_gauge_dependence} and \eqref{eq:redshift_gauge_dependence} are gauge dependent quantities representing the perturbations of redshift, incoming directions and the signal being on the past light-cone in a generic gauge. We can fix their gauge dependence by requiring that $\delta z$, $\delta w$ and $\delta\theta^a$ are null (i.e. these assume their measured values and are not perturbed). This gives
\be
B+E'= \frac{\delta z_{GI}}{\mathcal{H}}
\qquad,\qquad
\chi_r=-\delta w_{GI}-\frac{\delta z_{GI}}{\mathcal{H}}
\qquad\text{and}\qquad
\chi^a=-\delta\theta^a_{GI}\,.
\label{eq:observed_gauge}
\ee
Eqs.~\eqref{eq:observed_gauge} then correspond to identify the gauge invariant variables as the ones where redshift, light-cone and deflection angles are not perturbed. With this  it is straightforward to express $\delta\gamma^{ab}$ in terms of the choices \eqref{eq:observed_gauge}, and we obtain from Eq.~\eqref{eq:delta_gamma}
\bea
\delta\gamma^{ab}\left( z,{\bf n} \right)
&=&\delta\gamma^{ab}_{GI}
-2\,\bar\gamma^{ab}\,\delta z_{GI} ~ 
-\pa_r\bar\gamma^{ab}\delta w_{GI}
-\pa_r\bar\gamma^{ab}\frac{\delta z_{GI}}{\mathcal{H}}
-\delta\theta^c_{GI}\pa_c\bar\gamma^{ab}
\nonumber\\
&=&\delta\gamma^{ab}_{GI}
+2\,\bar\gamma^{ab}\,\left(
\frac{\delta z_{GI}}{\mathcal{H}r} -\delta z_{GI}
+\frac{\delta w_{GI}}{r}\right)
-\delta\theta^c_{GI}\pa_c\bar\gamma^{ab}\,,
\label{eq:delta_gamma_GI}
\eea
where we explicitly  note that Eq.~\eqref{eq:delta_gamma_GI} is the gauge invariant variable built in the gauge where redshift and light-cone perturbations are vanishing by construction\footnote{Another possibility to obtain the same result would have been by making a variable transform of $t$ into $z$ and $(x^i)$ into $n^i$ and the lightcone as usually done in literature \cite{Bonvin:2005ps,Fanizza:2015swa,Scaccabarozzi:2017ncm}. This gauge-invariant construction plays a crucial role also in the well-posedness of the light-cone averages of cosmological observables \cite{Gasperini:2011us,Yoo:2017svj,Fanizza:2019pfp,Buchert:2019mvq}.}.

Finally, thanks to the transformation properties of the Jacobian under the coordinate transformation from carthesian to polar coordinates and the fact that $H^{(t)}_{ij}$ is a tensor under $SO(3)$, we obtain that
\beq
\delta\gamma^{ab}_{GI}=\bar{\gamma}^{ac}\pa_c \delta\theta^b_{GI}
+\bar{\gamma}^{bc}\pa_c \delta\theta^a_{GI}
-2\,a^2\bar{r}^2\bar\gamma^{ac}\,H^{(t)}_{cd}\bar\gamma^{db}
+2\,\bar\gamma^{ab}\Psi\,,
\eeq
where we defined $H^{(t)}_{ab}=\pa_a n^i\pa_b n^j\, H^{(t)}_{ij}$.
The final expression is then
\bea
\delta\gamma^{ab}(z,{\bf n})&=&
\bar{\gamma}^{ac} \nabla_c\delta\theta^b_{GI}
+\bar{\gamma}^{bc} \nabla_c\delta\theta^a_{GI}
-2\,a^2\bar{r}^2\bar\gamma^{ac}\,H^{(t)}_{cd}\bar\gamma^{db}
\nonumber\\
&&+2\,\bar\gamma^{ab}\,\left(\Psi
-\delta z_{GI}
+\frac{\delta z_{GI}}{\mathcal{H}r}
+\frac{\delta w_{GI}}{r}\right)\,.
\label{eq:A25}
\eea
Hence, the Jacobi map on the past light-cone in terms of the observed redshift and the direction of observation is
\bea
J^A_B(z,{\bf n})
=\bar{d}(z,{\bf n})\left(\delta^A_B
+\frac{1}{2}\delta\hat{\gamma}_{ab}\hat{s}^b_A\hat{s}^a_B\right)\,,
\label{eq:A26}
\eea
where 
$\delta \gamma_{ab}=\bar{d}^{2}\delta\hat\gamma_{ab}$\,.
Combining Eqs.~\eqref{eq:A25} and \eqref{eq:A26} and recalling that in linear theory $\de\gamma_{ab}=-\bar\gamma_{ac}\de\gamma^{cd}\bar\gamma_{db}$, we find
\bea
J^A_B(z,{\bf n})&=&\bar d(z,{\bf n})
\left\{\delta_{AB}
-\frac{1}{2}\hat{\gamma}_{ac} \nabla_b\delta\theta^c_{GI}\hat{s}^b_A\hat{s}^a_B
-\frac{1}{2}\hat{\gamma}_{bc} \nabla_a\delta\theta^c_{GI}\hat{s}^b_A\hat{s}^a_B
+H^{(t)}_{ab}\hat{s}^b_A\hat{s}^a_B\right.\nonumber\\
&&\left.-\left[\Psi
-\left(1-\frac{1}{\mathcal{H}r}\right)\delta z_{GI}
+\frac{\delta w_{GI}}{r}\right]\delta_{AB}\right\}\,.
\eea

\section{Tensor power Spectra}
\subsection{Shear}\label{a:shear}
In this appendix we outline the procedure to compute the angular power spectrum of the shear from tensor perturbations given in \eqref{eq:ga_pm}. As the shear from scalar perturbations has been calculated in the literature before, we do not repeat it here.
We use the total angular momentum method and follow the procedure introduced, e.g., in Chapter 5 of Ref.~\cite{durrer_2020}.
In order to Fourier transform the terms appearing in \eqref{eq:ga_pm}, we first notice that $H^{(t)}_{rr}$, $H^{(t)}_{r\pm}$ and $H^{(t)}_{\pm\pm}$ are projections respectively of helicity $0$, $\pm1$ and $\pm2$ of the tensor perturbations in the basis $\{{\bf e}_\pm \}$ normal to the radial direction. Note that even though a gravitational wave has helicity 2 in the plane normal to its direction of propagation, it can carry helicities 0 and 1 in our direction of observation $\bn$. In total generality, the components $H_{ij}$  can then be decomposed as
\bea
H^{(t)}_{rr}(z,{\bf n})
&=&-\frac{4}{\sqrt{6}}\int\frac{d^3{\bf k}}{(2\pi)^3}\sum_{p=\pm 1}H_p(z,{\bf k})\,_0 G_{2\,2p}(z,{\bf n},\hat{{\bf k}})
\nonumber\\
H^{(t)}_{r\pm}(z,{\bf n})
&=&\pm\sqrt{2}\int\frac{d^3{\bf k}}{(2\pi)^3}\sum_{p=\pm 1}H_p(z,{\bf k})\,_{\pm1} G_{2\,2p}(z,{\bf n},\hat{{\bf k}})
\nonumber\\
H^{(t)}_{\pm\pm}(z,{\bf n})
&=&-2 \int\frac{d^3{\bf k}}{(2\pi)^3}\sum_{p=\pm 1}H_p(z,{\bf k})\,_{\pm2} G_{2\,2p}(z,{\bf n},\hat{{\bf k}})\,,
\label{eq:explicit_projections}
\eea
where we have introduced the basis functions
\be
\,_sG_{\ell m}(z,{\bf n},\hat{{\bf k}})\equiv(-i)^\ell\sqrt{\frac{4\pi}{2\ell +1}}\,_sY_{\ell m}({\bf n};\hat{{\bf k}})e^{i{\bf k}\cdot{\bf n}}\,.
\ee
Here $_sY_{\ell m}({\bf n},\hat{{\bf k}})$ is the spherical harmonic $_sY_{\ell m}({\bf n})$ where $\bf k$ is identified as the ${\bf e}_z$ direction, i.e., the angles of ${\bf n}$ evaluated w.r.t.~the direction of ${\bf k}$, see~\cite{durrer_2020} for details.
The exponential $e^{i{\bf k}\cdot{\bf n}}$ can be further expanded into spherical Bessel functions and spherical harmonics so that the total angular momentum decomposition leads to~\cite{durrer_2020}
\be
\,_sG_{\ell m}(z,{\bf n},\hat{{\bf k}})=\sum_{j=0}^{+\infty}\sqrt{4\pi\left( 2j+1 \right)}\left( -i \right)^j\,_sf^{\ell m}_j(k r)\,_sY_{j m}({\bf n};\hat{{\bf k}})\,,
\label{eq:basis}
\ee
where 
\be
\,_s f^{\ell m}_j(x)=\sum_{L=|j-\ell|}^{j+\ell}(-i)^{L+\ell-j}\frac{2L+1}{2j+1}
\langle L,\ell;0,m|j,m\rangle
\langle L,\ell;0,-s|j,-s\rangle\, 
j_L(x) \,.
\label{eq:cg}
\ee
Here $j_L(x)$ is the spherical Bessel function of $L$-th order and $\langle \ell_1,\ell_2;m_1,m_2|\ell_3,m_3 \rangle$ are the Clebsch-Gordan coefficients. All these functions are  used with the conventions reported in \cite{durrer_2020}.

The coefficients in front of the integrals in Eqs.~\eqref{eq:explicit_projections} are needed in order to obtain the correct limit when ${\bf k}\;\rVert\; {\bf e}_z$ where the polarization tensors are
\be
e^{\pm}_{ij}(\hat{\bf k})=\left(
\begin{matrix}
\sigma_3 \pm \,i\sigma_1&\vec{0}\\
\vec{0}^{\,T}&0
\end{matrix}\right)
\ee
where $\sigma_3$ and $\sigma_1$ are Pauli matrices, such that e.g.
(see also~\cite{Schmidt:2012ne})
\bea
\epsilon^p_{ij}(\hat{\bf k})e^i_{\pm} e^j_{pm}&=&
\frac{1}{2}\left( 1\mp p\cos\theta \right)^2e^{2ip\phi}\,,
\nonumber\\
\epsilon^p_{ij}(\hat{\bf k})n^i n^j&=&
\left( 1-\cos^2\theta \right)e^{2ip\phi}\,,
\nonumber\\
\epsilon^p_{ij}(\hat{\bf k})e^i_{\pm} n^j&=&
\sqrt{\frac{1-\cos^2\theta}{2}}\left( \cos\theta\mp p\right)e^{2ip\phi}\,,
\eea
where $e^p_{ij}(\hat{\bf k})$ is the transverse, traceless polarization tensor and $p=\pm 1$ accounts for the two polarization states of the tensor perturbations. Before continuing, a comment about the orientation of the basis is in order. Indeed, given the tensor nature of $H^{(t)}_{ij}$, we need to care about the orientation of the basis and, in particular, we must ensure that the final result is independent of this choice. Given that, the choice of ${\bf k}$ as a reference basis might look peculiar and it is certainly not well suited in light of the $\int d^3{\bf k}$ integrals in Eqs.~\eqref{eq:explicit_projections}. To overcome this issue, we need to refer to an generic basis ${\bf E}$ and decompose our perturbations in terms of this reference basis. This change of reference frame leads to (see \cite{durrer_2020})
\be
\,_sY_{\ell\,m}({\bf n};\hat{{\bf k}})=\sqrt{\frac{4\pi}{2\ell+1}}\sum_{j} \,_sY_{\ell j}({\bf n};{\bf E})\,_{-m}Y^*_{\ell j}(\hat{{\bf k}};{\bf E})\,.
\label{eq:rotation}
\ee

In this way, Eqs.~\eqref{eq:explicit_projections} are ready to be manipulated. With Eqs.~\eqref{eq:basis} and \eqref{eq:rotation} we obtain

\bea
\pa^2_\mp\pa^2_\pm H^{(t)}_{rr}(z,{\bf n})
&=&-\frac{4}{\sqrt{6}}\int\frac{d^3{\bf k}}{(2\pi)^3}\sum_{p=\pm 1}\sum_{j=0}^{+\infty}H_p(z,{\bf k})
\nonumber\\
&&\times\sqrt{4\pi\left( 2j+1 \right)}\left( -i \right)^j\,_0f^{2\,2p}_j(k r)\pa^2_\mp\pa^2_\pm \left[Y_{j\,2p}({\bf n};\hat{{\bf k}})\right]
\nonumber\\
&=&-\frac{4\pi}{\sqrt{6}}\int\frac{d^3{\bf k}}{(2\pi)^3}\sum_{p=\pm 1}\sum_{j=0}^{+\infty}
\nonumber\\
&&\times
\sum_{m}\left( -i \right)^j\,_0f^{2\,2p}_j(k r)\frac{(j+2)!}{(j-2)!}
Y_{jm}({\bf n};{\bf E})\,_{-2p}Y^*_{jm}(\hat{{\bf k}};{\bf E})H_p(z,{\bf k})\,,
\nonumber\\
\pa^2_\mp\pa_\pm H^{(t)}_{r\pm}(z,{\bf n})
&=&\pm\sqrt{2}\int\frac{d^3{\bf k}}{(2\pi)^3}\sum_{p=\pm 1}\sum_{j=0}^{+\infty}H_p(z,{\bf k})
\nonumber\\
&&\times\sqrt{4\pi\left( 2j+1 \right)}\left( -i \right)^j\,_{\pm 1}f^{2\,2p}_j(k r)\pa^2_\mp\pa_\pm\left[\,_{\pm 1}Y_{j\,2p}({\bf n};\hat{{\bf k}})\right]
\nonumber\\
&=&-\frac{4\pi}{2}\int\frac{d^3{\bf k}}{(2\pi)^3}\sum_{p=\pm 1}\sum_{j=0}^{+\infty}
\nonumber\\
&&\times
\sum_{m}\left( -i \right)^j\,_{\pm 1}f^{2\,2p}_j(k r)\frac{(j+2)!}{(j-2)!}\sqrt{\frac{(j-1)!}{(j+1)!}}
Y_{jm}({\bf n};{\bf E})\,_{-2p}Y^*_{jm}(\hat{{\bf k}};{\bf E})H_p(z,{\bf k})\,,
\nonumber\\
\pa^2_\mp H^{(t)}_{\pm\pm}(z,{\bf n})
&=&-2\int\frac{d^3{\bf k}}{(2\pi)^3}\sum_{p=\pm 1}\sum_{j=0}^{+\infty}H_p(z,{\bf k})
\nonumber\\
&&\times\sqrt{4\pi\left( 2j+1 \right)}\left( -i \right)^j\,_{\pm 2}f^{2\,2p}_j(k r)\pa^2_{\mp}\left[\,_{\pm 2}Y_{j\,2p}({\bf n};\hat{{\bf k}})\right]
\nonumber\\
&=&-4\pi\int\frac{d^3{\bf k}}{(2\pi)^3}\sum_{p=\pm 1}\sum_{j=0}^{+\infty}
\nonumber\\
&&\times
\sum_{m}\left( -i \right)^j\,_{\pm 2}f^{2\,2p}_j(k r)\sqrt{\frac{(j+2)!}{(j-2)!}}
Y_{jm}({\bf n};{\bf E})\,_{-2p}Y^*_{jm}(\hat{{\bf k}};{\bf E})H_p(z,{\bf k})\,.
\label{eq:explicit_projections_2}
\eea

With the help of Eqs.~\eqref{eq:explicit_projections_2}, we can now explicitly evaluate the $a^\pm_{\ell m}$'s in Eq.~\eqref{eq:decomposition}, where the direction ${\bf n}$ in the definition of the spherical harmonics refers to a general basis ${\bf E}$ and we are sure that the result is independent of this choice. In order to evaluate the $a^{E/B}_{\ell m}(z)$, we first recall that $\ds=-\sqrt{2}\pa_+$ and $\bds=-\sqrt{2}\pa_-$ and we then have
\bea
\bds^2\ga_+\left(z,{\bf n}\right)&=&\sum_{\ell m}a^+_{\ell m}(z)\,\bds^2\,_{+ 2}Y_{\ell m}({\bf n};{\bf E})
=\sum_{\ell m}a^+_{\ell m}(z)\,\sqrt{\frac{\left( \ell+2 \right)!}{\left( \ell-2 \right)!}}\,Y_{\ell m}({\bf n};{\bf E})\,,
\nonumber\\
\ds^2\ga_-\left(z,{\bf n}\right)&=&\sum_{\ell m}a^-_{\ell m}(z)\,\ds^2_{- 2}Y_{\ell m}({\bf n};{\bf E})
=\sum_{\ell m}a^-_{\ell m}(z)\,\sqrt{\frac{\left( \ell+2 \right)!}{\left( \ell-2 \right)!}}\,Y_{\ell m}({\bf n};{\bf E})\,,
\eea
leading to
\bea
a^+_{\ell m}(z)&=&\sqrt{\frac{\left( \ell-2 \right)!}{\left( \ell+2 \right)!}}\int d\Omega_{\bf n}\,\bds^2\ga_+\left(z,{\bf n}\right)Y^*_{\ell m}({\bf n};{\bf E})\,,
\nonumber\\
a^-_{\ell m}(z)&=&\sqrt{\frac{\left( \ell-2 \right)!}{\left( \ell+2 \right)!}}\int d\Omega_{\bf n}\,\ds^2\ga_-\left(z,{\bf n}\right)Y^*_{\ell m}({\bf n};{\bf E})\,.
\eea
With this the $a^{E/B}_{\ell m}(z)$ coefficients in Eq.~\eqref{eq:coeff} can be readily expressed as
\be
a^\pm_{\ell m}(z_s)=\int_0^{r_s}dr\frac{r_s-r}{r\,r_s}\Scal^\pm_{\ell m}(z)
+\int_0^{r_s}\frac{dr}{r}\Vcal^\pm_{\ell m}(z)
+\Tcal^\pm_{\ell m}(z_s)\,,
\label{eq:aSVT}
\ee
where we  define respectively the helicity 0, 1 and 2 projections of the tensor perturbations onto the line-of-sight as
\bea
\Scal^\pm_{\ell m}(z)
&=&-\left( -i \right)^\ell\frac{8\pi}{\sqrt{6}}\sqrt{\frac{(\ell+2)!}{(\ell-2)!}}\int\frac{d^3{\bf k}}{(2\pi)^3}\sum_{p=\pm 1}
\,_0f^{2\,2p}_\ell(k r)
\,_{-2p}Y^*_{\ell m}(\hat{{\bf k}};{\bf E})H_p(z,{\bf k})\,,
\nonumber\\
\Vcal^\pm_{\ell m}(z)
&=&\left( -i \right)^\ell 8\pi \sqrt{\frac{(\ell+2)!}{(\ell-2)!}\frac{(\ell-1)!}{(\ell+1)!}}\int\frac{d^3{\bf k}}{(2\pi)^3}\sum_{p=\pm 1}
\,_{\pm 1}f^{2\,2p}_\ell(k r)
\,_{-2p}Y^*_{\ell m}(\hat{{\bf k}};{\bf E})H_p(z,{\bf k})\,,
\nonumber\\
\Tcal^\pm_{\ell m}(z)
&=&-\left( -i \right)^\ell 8\pi\int\frac{d^3{\bf k}}{(2\pi)^3}\sum_{p=\pm 1}
\,_{\pm 2}f^{2\,2p}_\ell(k r)
\,_{-2p}Y^*_{\ell m}(\hat{{\bf k}};{\bf E})H_p(z,{\bf k})\,.
\label{eq:alm_SVT}
\eea
Here {\bf E} is an arbitrary fixed direction and
\bea
\,_0 f^{2\,2p}_\ell(x)&=&\sqrt{\frac{3(\ell+2)!}{8(\ell-2)!}}\frac{j_\ell(x)}{x^2}\,,
\nonumber\\
\,_{\pm 1} f^{2\,2p}_\ell(x)
&=&\frac{1}{2x}\sqrt{\frac{(\ell-2)!(\ell+1)!}{(\ell+2)!(\ell-1)!}}
\left\{ (\ell-1)(\ell+2)\left[ (\ell+1)\frac{j_\ell(x)}{x}-j_{\ell+1}(x)\right]\right.
\nonumber\\
&&\left.\mp i\,p\left[ (\ell^2-\ell-3)j_\ell (x)+x\left[ j_{\ell-1}(x)+j_{\ell+1}(x) \right] \right]\right\}\,,
\nonumber\\
\,_{\pm 2} f^{2\,2p}_\ell(x)
&=&\frac{\left[ \ell(\ell -1) -2 x^2\right] j_{\ell }(x)+2 x j_{\ell -1}(x)}{4 x^2}
\mp \frac{i\,p}{2}\left[(\ell +2)\frac{ j_{\ell }(x)}{x}- j_{\ell +1}(x)\right]\,. \qquad
\label{eq:fs}
\eea
Let us underline that Eqs.~\eqref{eq:fs} directly follow from Eq.~\eqref{eq:cg}. 

We are now in the position to evaluate the $E$- and $B$- modes  by following the decomposition in Eq.~\eqref{eq:coeff}. We keep the helicity classification and find
\bea
\Scal^E_{\ell m}(z)
&=&-\left( -i \right)^\ell 4\pi\frac{(\ell+2)!}{(\ell-2)!}\int\frac{d^3{\bf k}}{(2\pi)^3}
\frac{1}{2}\frac{j_\ell(kr)}{(kr)^2}
\sum_{p=\pm 1}\,_{-2p}Y^*_{\ell m}(\hat{{\bf k}};{\bf E})H_p(z,{\bf k})\,,
\nonumber\\
\Scal^B_{\ell m}(z)
&=&0 \,,
\nonumber
\eea
\bea
\Vcal^E_{\ell m}(z)
&=&-\left( -i \right)^\ell 4\pi (\ell-1)(\ell+2)\int\frac{d^3{\bf k}}{(2\pi)^3}
\left[ \frac{j_{\ell+1}(kr)}{kr}-(\ell+1)\frac{j_\ell(kr)}{(kr)^2}\right]
\nonumber\\
&&\times\sum_{p=\pm 1}\,_{-2p}Y^*_{\ell m}(\hat{{\bf k}};{\bf E})H_p(z,{\bf k})\,,
\nonumber\\
\Vcal^B_{\ell m}(z)
&=&\left( -i \right)^\ell 4\pi \int\frac{d^3{\bf k}}{(2\pi)^3}
\left[ (\ell^2-\ell-3)\frac{j_\ell (kr)}{kr}+ j_{\ell-1}(kr)+j_{\ell+1}(kr) \right]
\nonumber\\
&&\times\sum_{p=\pm 1}
\,p\,_{-2p}Y^*_{\ell m}(\hat{{\bf k}};{\bf E})H_p(z,{\bf k})\,,
\\
\Tcal^E_{\ell m}(z)
&=&-\left( -i \right)^\ell 4\pi\int\frac{d^3{\bf k}}{(2\pi)^3}
\left[\frac{\ell(\ell -1)}{2} \frac{j_{\ell }(kr)}{(kr)^2}
-j_{\ell }(kr)+\frac{j_{\ell -1}(kr)}{kr}\right]
\nonumber\\
&&\sum_{p=\pm 1}\,_{-2p}Y^*_{\ell m}(\hat{{\bf k}};{\bf E})H_p(z,{\bf k})\,,
\nonumber\\
\Tcal^B_{\ell m}(z)
&=&-\left( -i \right)^\ell 4\pi\int\frac{d^3{\bf k}}{(2\pi)^3}\left[(\ell +2)\frac{ j_{\ell}(kr)}{kr}-j_{\ell +1}(kr)\right]
\sum_{p=\pm 1}p\,_{-2p}Y^*_{\ell m}(\hat{{\bf k}};{\bf E})H_p(z,{\bf k})
\,.
\nonumber
\label{eq:EB_SVT}
\eea
In order to link these  $a_{\ell m}$'s  to the $C_\ell$'s, we invoke the statistical isotropy and homogeneity of the {\it ensemble average} (the power spectrum depends only on the absolute value $k$ but not on directions)
\bea
\langle H_{p}(z_1,{\bf k})H^*_n(z_2,{\bf q}) \rangle
&=&(2\pi)^3\delta({\bf k}-{\bf q})\delta_{p n}\frac{P_h(k,z_1,z_2)}{4}
\nonumber\\
&=&(2\pi)^3\delta({\bf k}-{\bf q})\delta_{p n}\frac{P_T(k,z_1,z_2)}{32}\,,
\label{eq:331}
\eea

A comment about the numerical factor in Eq.~\eqref{eq:331} is in order. Indeed, we first decompose $H^{(t)}_{ij}=e^+_{ij}H_++e^\times_{ij} H_\times$, where
\begin{equation}
\langle H_+(z_1,{\bf k})H^*_+(z_2,{\bf q}) \rangle
=\langle H_\times(z_1,{\bf k})H^*_\times(z_2,{\bf q}) \rangle
=(2\pi)^3\delta({\bf k}-{\bf q})\frac{P_T(k,z_1,z_2)}{4}\,.
\end{equation}
Factor 4 is needed in order to have
\begin{equation}
\langle H_{ij}(z_1,{\bf k})H^{ij*}(z_1,{\bf q}) \rangle
=(2\pi)^3\delta({\bf k}-{\bf q})P_T(k,z_1,z_2)\,,
\end{equation}
since $e^+_{ij}e^{+ij}=e^\times_{ij}e^{\times ij}=2$.
In Eq.~\eqref{eq:explicit_projections} instead, we express the tensor perturbations in terms of
\begin{equation}
H_{\pm 1}=\frac{1}{2}\left( H_+\mp i H_\times \right)\,,
\end{equation}
hence
\begin{align}
\langle H_1(z_1,{\bf k})H^*_1(z_2,{\bf q}) \rangle
=&\langle H_{-1}(z_1,{\bf k})H^*_{-1}(z_2,{\bf q}) \rangle
\nonumber\\
=&\frac{1}{4}\left( \langle H_+(z_1,{\bf k})H^*_+(z_2,{\bf q}) \rangle
+\langle H_\times(z_1,{\bf k})H^*_\times(z_2,{\bf q}) \rangle\right)
\nonumber\\
=&(2\pi)^3\delta({\bf k}-{\bf q})\frac{P_T(k,z_1,z_2)}{8}\,.
\end{align}
This derivation is in line with the convention of \cite{Schmidt:2012ne}. However, a further factor 4 is needed to take into account our factor 2 in the metric definition \eqref{e:met1}. By taking this into account, we arrive at Eq.~\eqref{eq:331}.
The ensemble average is needed since it enters in the definition of the multipoles,
\be
C^{XY}_\ell(z_1,z_2)=\langle a^{X}_{\ell m}(z_1)a^{Y*}_{\ell m}(z_2)\rangle\,,
\label{eq:Cls}
\ee
where $X,Y=E,B$. We notice that whenever we apply the definition \eqref{eq:Cls} to any pair of terms in Eqs.~\eqref{eq:EB_SVT}, we always obtain the  following structure
\bea
&&\int\frac{d^3{\bf k}}{(2\pi)^3}
\int\frac{d^3{\bf q}}{(2\pi)^3}F(k)G(q)
\sum_{p,n=\pm 1}p^\alpha n^\beta
\,_{-2p}Y^*_{\ell m}(\hat{{\bf k}};{\bf E})\,_{-2n}Y_{\ell' m'}(\hat{\bf q};{\bf E})
\langle H_p(z_1,{\bf k})H^*_n(z_2,{\bf q})\rangle
\nonumber\\
&=&\sum_{p=\pm 1}p^{\alpha+\beta}\int\frac{d^3{\bf k}}{(2\pi)^3}F(k)G(k)
\,_{-2p}Y^*_{\ell m}(\hat{{\bf k}};{\bf E})\,_{-2p}Y_{\ell' m'}(\hat{{\bf k}};{\bf E})
P_h(k,z_1,z_2)
\nonumber\\
&=& N_{\alpha\beta}\,\delta_{\ell\ell'}\delta_{mm'}
\int\frac{k^2dk}{(2\pi)^3}F(k)G(k)
P_h(k,z_1,z_2)\,,
\eea
where $F$ and $G$ are two unspecified functions depending only on the modulus of the wave vector and we have used the orthonormality of the $\,_s Y_{\ell m}(\hat{\bf k};{\bf E})$ in Eq.~\eqref{eq:D3}. We define the factor $N_{\alpha\beta}\equiv\sum_{p=\pm 1}p^{\alpha+\beta}$, where $\alpha\,,\,\beta=0$ for $E$-modes and 1 for $B$-modes. From Eqs.~\eqref{eq:EB_SVT} we notice that the sum $\alpha+\beta$ is $0$ or $2$ respectively for $EE$ and $BB$ correlations, whereas it is $1$ for $EB$ correlations. This implies that $N_{\alpha\beta} =2$ for both $EE$ and $BB$ modes whereas we get $N_{\alpha\beta}=0$ for $EB$ modes. This is a  consequence of the fact that $E$- and $B$- modes have opposite parity which implies that their correlation vanishes.

We can now write the non-vanishing terms appearing in the $C_\ell$'s of the shear $E$- and $B$- modes. In particular, we split
\bea
C^{X}_\ell(z_1,z_2)&\equiv&\,_0C^{X}(z_1,z_2)+\,_1C^{X}(z_1,z_2)+\,_2C^{X}(z_1,z_2)
\nonumber\\
&&+\,_{01}C^{X}(z_1,z_2)+\,_{02}C^{X}(z_1,z_2)+\,_{12}C^{X}(z_1,z_2)\,,
\eea
where $0,1,2$ respectively label helicity 0 ($\Scal$), helicity 1 ($\Vcal$) and helicity 2 ($\Tcal$) projections of the tensor modes. Single subscript stands for auto-correlation of spin projections whereas double subscripts indicates cross-correlations between different spin projections. We {obtain the following results for the non-vanishing} auto-correlations:
\bea
\,_0C^E_\ell(z_1,z_2)
&=&\frac{1}{32\pi}\left[\frac{(\ell+2)!}{(\ell-2)!}\right]^2
\int_0^{r_1}dr\int_0^{r_2}dr'\frac{r_1-r}{r r_1}\frac{r_2-r'}{r' r_2}
\nonumber\\
&&\times
\int k^2dk P_T(k,z,z')
\frac{j_\ell(kr)}{(kr)^2}
\frac{j_\ell(kr')}{(kr')^2}\,,
\nonumber\\
\,_1C^E_\ell(z_1,z_2)
&=&\frac{(\ell-1)^2(\ell+2)^2}{8\pi}
\int_0^{r_1}\frac{dr}{r}\int_0^{r_2}\frac{dr'}{r'}
\int k^2dk P_T(k,z,z')
\nonumber\\
&&\times
\left[ \frac{j_{\ell+1}(kr)}{kr}-(\ell+1)\frac{j_\ell(kr)}{(kr)^2}\right]
\left[ \frac{j_{\ell+1}(kr')}{kr'}-(\ell+1)\frac{j_\ell(kr')}{(kr')^2}\right]\,,
\nonumber\\
\,_2C^E_\ell(z_1,z_2)
&=&\frac{1}{8\pi}
\int k^2dk P_T(k,z_1,z_2)
\left[\frac{\ell(\ell -1)}{2} \frac{j_{\ell }(kr_1)}{(kr_1)^2}
-j_{\ell }(kr_1)+\frac{j_{\ell -1}(kr_1)}{kr_1}\right]
\nonumber\\
&&\times\left[\frac{\ell(\ell -1)}{2} \frac{j_{\ell }(kr_2)}{(kr_2)^2}
-j_{\ell }(kr_2)+\frac{j_{\ell -1}(kr_2)}{kr_2}\right]\,,
\nonumber
\eea
\bea
\,_1C^B_\ell(z_1,z_2)
&=&\frac{1}{8\pi}
\int_0^{r_1}\frac{dr}{r}\int_0^{r_2}\frac{dr'}{r'}
\int k^2dk P_T(k,z,z')
\nonumber\\
&&\times\left[ (\ell^2-\ell-3)\frac{j_\ell (kr)}{kr}+ j_{\ell-1}(kr)+j_{\ell+1}(kr) \right]
\nonumber\\
&&\times\left[ (\ell^2-\ell-3)\frac{j_\ell (kr')}{kr'}+ j_{\ell-1}(kr')+j_{\ell+1}(kr') \right]\,,
\nonumber\\
\,_2C^B_\ell(z_1,z_2)
&=&\frac{1}{8\pi}
\int k^2dk P_T(k,z_1,z_2)
\nonumber\\
&&\times\left[(\ell +2)\frac{ j_{\ell}(kr_1)}{kr_1}-j_{\ell +1}(kr_1)\right]
\left[(\ell +2)\frac{ j_{\ell}(kr_2)}{kr_2}-j_{\ell +1}(kr_2)\right]\,,
\label{eqA:auto}
\eea
whereas the non-vanishing cross-correlations are
\bea
\,_{01}C^E_\ell(z_1,z_2)
&=&\frac{(\ell-1)(\ell+2)}{16\pi}\frac{(\ell+2)!}{(\ell-2)!}\int_0^{r_1}dr\frac{r_1-r}{rr_1}\int_0^{r_2}\frac{dr'}{r'}
\int k^2dk P_T(k,z,z')
\nonumber\\
&&\times
\frac{j_\ell(kr)}{(kr)^2}
\left[ \frac{j_{\ell+1}(kr')}{kr'}-(\ell+1)\frac{j_\ell(kr')}{(kr')^2}\right]
+\left( z_1 \leftrightarrow z_2 \right)\,,
\nonumber\\
\,_{02}C^E_\ell(z_1,z_2)
&=&\frac{1}{16\pi}\frac{(\ell+2)!}{(\ell-2)!}\int_0^{r_1}dr\frac{r_1-r}{rr_1}
\int k^2dk P_T(k,z,z_2)
\nonumber\\
&&\times\frac{j_\ell(kr)}{(kr)^2}
\left[\frac{\ell(\ell -1)}{2} \frac{j_{\ell }(kr_2)}{(kr_2)^2}
-j_{\ell }(kr_2)+\frac{j_{\ell -1}(kr_2)}{kr_2}\right]
+\left( z_1 \leftrightarrow z_2 \right)\,,
\nonumber\\
\,_{12}C^E_\ell(z_1,z_2)
&=&\frac{(\ell-1)(\ell+2)}{8\pi}\int_0^{r_1}\frac{dr}{r}
\int k^2dk P_T(k,z,z_2)
\left[ \frac{j_{\ell+1}(kr)}{kr}-(\ell+1)\frac{j_\ell(kr)}{(kr)^2}\right]
\nonumber\\
&&\times\left[\frac{\ell(\ell -1)}{2} \frac{j_{\ell }(kr_2)}{(kr_2)^2}
-j_{\ell }(kr_2)+\frac{j_{\ell -1}(kr_2)}{kr_2}\right]
+\left( z_1 \leftrightarrow z_2 \right)\,,
\nonumber\\
\,_{12}C^B_\ell(z_1,z_2)
&=&-\frac{1}{8\pi}\int_0^{r_1}\frac{dr}{r}
\int k^2dk P_T(k,z,z_2)
\left[ (\ell^2-\ell-3)\frac{j_\ell (kr)}{kr}+ j_{\ell-1}(kr)+j_{\ell+1}(kr) \right]
\nonumber\\
&&\times\left[(\ell +2)\frac{ j_{\ell}(kr_2)}{kr_2}-j_{\ell +1}(kr_2)\right]
+\left( z_1 \leftrightarrow z_2 \right)\,.
\label{eqA:cross}
\eea

\subsection{Comparison with Schmidt and Jeong
\label{as:comp}}

The total results derived  in this work nicely agree with the evaluation of the $E$- and $B$- modes for the shear lensing in \cite{Schmidt:2012ne,Schmidt:2012nw}. However, our classification of the involved terms is different. Indeed, in \cite{Schmidt:2012ne,Schmidt:2012nw} the $a^\pm_{\ell m}$ coefficients have been split in the following way
\bea
a^+_{\ell m}(z_s)&\sim&\int_0^{r_s}\frac{dr}{r}\left[ \widehat{Q}_2(kr)+\frac{r}{r_s}\widehat{Q}_3(kr)  \right]\frac{j_\ell(kr)}{(kr)^2}
-\frac{1}{4}\widehat{Q}_1(kr_s)\frac{j_\ell(kr_s)}{(kr_s)^2}\,,
\nonumber\\
a^-_{\ell m}(z_s)&\sim&\int_0^{r_s}\frac{dr}{r}\left[ \widehat{Q}^*_2(kr)+\frac{r}{r_s}\widehat{Q}^*_3(kr)  \right]\frac{j_\ell(kr)}{(kr)^2}
-\frac{1}{4}\widehat{Q}^*_1(kr_s)\frac{j_\ell(kr_s)}{(kr_s)^2}\,,
\label{eq:SJ}
\eea
where $\widehat{Q}_1$, $\widehat{Q}_2$ and $\widehat{Q}_3$ are complex operators acting on $j_\ell(x)/x^2$ function. The explicit definition of these operators can be found in the Appendix B of \cite{Schmidt:2012nw} and the total sums in the combinations of $a^\pm_{\ell m}$ agree with our results. The direct comparison between Eq.~\eqref{eq:SJ} and \eqref{eq:alm_SVT} yields the relations
\be
\widehat{Q}_1\propto-\Tcal
\qquad,\qquad
\widehat{Q}_2\propto\Scal+\Vcal
\qquad\text{and}\qquad
\widehat{Q}_3\propto-\Scal\,,
\label{eq:B26}
\ee
where the notation $\propto$ here is meant to be valid once the operators in Eq.~\eqref{eq:B26} are applied on $j_\ell(x)/x^2$.

The helicity-2 projections are decoupled from the others in both ways of splitting $\ga_\pm$.
 However, in Eq.~\eqref{eq:SJ} the helicity-0 and helicity-1 projections of  tensor perturbations are not well-decoupled. In particular, helicity-0 projection sources both $\widehat{Q}_2$ and $\widehat{Q}_3$. For the $B$- mode, where $\Scal^B=0$, both the procedures lead to a proper decoupling between helicity-1 and helicity-2 projections. On the contrary, for the $E$- mode the helicity-0 projection sources both $\widehat{Q}_2$ and $\widehat{Q}_3$.

\subsection{Convergence}\label{a:conv}
To compute the convergence power spectrum from tensor perturbations we  apply the same decomposition as in Eq.~\eqref{eq:explicit_projections_2}, we have 
\be
H^{(t)}_{rr}(z,{\bf n})
=-\frac{16\pi}{\sqrt{6}}\int\frac{d^3{\bf k}}{(2\pi)^3}\sum_{p=\pm 1}\sum_{j,m}
\left( -i \right)^j\,_0f^{2\,2p}_j(k r)
Y_{jm}({\bf n};{\bf E})\,_{-2p}Y^*_{jm}(\hat{{\bf k}};{\bf E})H_p(z,{\bf k})\,.
\label{eq:44}
\ee
At this point, we just need to evaluate the radial and angular derivatives of Eq.~\eqref{eq:44}, namely we need the following quantities
\bea
\pa_r\left[\,_0f_\ell^{2\,2p}(kr)\right]
&=&\frac{1}{r}\sqrt{\frac{3(\ell+2)!}{8(\ell-2)!}}
\left[(\ell -2) \frac{j_{\ell }(kr)}{(kr)^2}-\frac{j_{\ell +1}(kr)}{kr}\right]\,,
\nonumber\\
\Delta_2 Y_{\ell m}({\bf n};{\bf E})&=&-\ell(\ell+1)Y_{\ell m}({\bf n};{\bf E})\,.
\label{eq:45}
\eea
Once we combine Eqs.~\eqref{eq:kappa_tensor_2}, \eqref{eq:44} and \eqref{eq:45}, we can readily evaluate the $a_{\ell m}$'s of $\kappa$ by noting that the latter is a scalar so that its harmonic coefficients are given by
\be
a^\kappa_{\ell m}(z)=\int d\Omega_{\bf n} \kappa(z,{\bf n}) Y^*_{\ell m}({\bf n};{\bf E})\,,
\ee
leading to
\bea
a^\kappa_{\ell m}(z)
&=&\int d\Omega_{\bf n}\kappa(z,{\bf n})Y^*_{\ell m}({\bf n};{\bf E})
\nonumber\\
&=&
-\frac{16\pi}{\sqrt{6}}\sum_{p=\pm 1}\int\frac{d^3{\bf k}}{(2\pi)^3}
\left( -i \right)^\ell
\left\{\int_0^{r}dr'\,
\left[
\left(\frac{1}{r}-\frac{3}{r'}\right)\,_0f_\ell^{2\,2p}(kr')\right.\right.
\nonumber\\
&&\left.\left.
-\left(2-\frac{1}{\mathcal{H}r}\right)\pa_{r'}\left[\,_0f_\ell^{2\,2p}(kr')\right]
+\frac{\ell(\ell+1)}{2} 
\frac{r-r'}{rr'}\,_0f_\ell^{2\,2p}(kr')\right]\,H_p(z',{\bf k})\right.
\nonumber\\
&&\left.+\left(\frac{3}{2}-\frac{1}{\mathcal{H}r}\right)\,_0f_\ell^{2\,2p}(kr) H_p(z,{\bf k})\right\}
\,_{-2p}Y^*_{\ell m}(\hat{{\bf k}};{\bf E})
\nonumber\\
&=&
-\left( -i \right)^\ell 4\pi\,\sqrt{\frac{(\ell+2)!}{(\ell-2)!}}\sum_{p=\pm 1}\int\frac{d^3{\bf k}}{(2\pi)^3}
\left\{\int_0^{r}\frac{dr'}{r'}\,
\left[\left(2-\frac{1}{\mathcal{H}r}\right)
\frac{j_{\ell +1}(kr')}{kr'}
\right.\right.
\nonumber\\
&&\left.\left.
+\left(
\frac{2-\ell(\ell+1)}{2}\frac{r'}{r}
-\frac{4(\ell -2)-\ell(\ell+1)}{2}
+\frac{\ell -2}{\mathcal{H}r}
-3\right)\frac{j_{\ell }(kr')}{(kr')^2}
\right]\,H_p(z',{\bf k})\right.
\nonumber\\
&&\left.+\left(\frac{3}{2}-\frac{1}{\mathcal{H}r}\right)
\frac{j_{\ell }(kr)}{(kr)^2} H_p(z,{\bf k})\right\}
\,_{-2p}Y^*_{\ell m}(\hat{{\bf k}};{\bf E})
\nonumber\\
&=&
-\left( -i \right)^\ell 4\pi\,\sqrt{\frac{(\ell+2)!}{(\ell-2)!}}\sum_{p=\pm 1}\int\frac{d^3{\bf k}}{(2\pi)^3}
\left[\int_0^{r}\frac{dr'}{r'}\Ical_\ell(k,r,r')\,
\,H_p(z',{\bf k})\right.
\nonumber\\
&&\left.+\Lcal_\ell(k,r) H_p(z,{\bf k})\right]
\,_{-2p}Y^*_{\ell m}(\hat{{\bf k}};{\bf E})\,,
\eea
where $r$ and $r'$ are the comoving distances to reshifts $z$ and $z'$ respectively and 
we have defined
\bea
\Ical_\ell(k,r,r')&=&\left(2-\frac{1}{\mathcal{H}r}\right)
\frac{j_{\ell +1}(kr')}{kr'}
\nonumber\\
&&+\left(
\frac{2-\ell(\ell+1)}{2}\frac{r'}{r}
-\frac{4(\ell -2)-\ell(\ell+1)}{2}
+\frac{\ell -2}{\mathcal{H}r}
-3\right)\frac{j_{\ell }(kr')}{(kr')^2}\,,
\nonumber\\
\Lcal_\ell(k,r)&=&\left(\frac{3}{2}-\frac{1}{\mathcal{H}r}\right)
\frac{j_{\ell }(kr)}{(kr)^2}\,.
\label{eq:kappa_kernels}
\eea
The corresponding kernels $\Ical^E_\ell$ and $\Lcal^E_\ell$ for the shear $E$-modes are
\bea
\Ical^E_\ell(k,r,r')&=&
\left[\frac{\ell(\ell+1)}{2}\frac{r-r'}{r}
-(\ell -2)
-3\right]\frac{j_{\ell }(kr')}{(kr')^2}
+\frac{j_{\ell +1}(kr')}{kr'}\,,
\nonumber\\
\Lcal^E_\ell(k,r)&=&\frac{1}{2}
\frac{j_{\ell }(kr)}{(kr)^2}\,.
\label{eq:gammaE_kernels}
\eea
A direct comparison between Eqs.~\eqref{eq:kappa_kernels} and \eqref{eq:gammaE_kernels}  shows that there is no simple
exact relation between the angular spectra of convergence and shear $E$- modes for tensor perturbations.

Again, with the help  of Eqs.~\eqref{eq:331} and \eqref{eq:Cls}, we obtain the angular spectrum of the convergence,
\bea
C^\kappa_\ell(z_1,z_2)
&=&\frac{1}{8\pi}\frac{(\ell+2)!}{(\ell-2)!}
\int k^2 dk\left\{
\int^{r_1}_0\frac{dr'}{r'}
\int^{r_2}_0\frac{dr''}{r''}P_T(k,z',z'')
\Ical_\ell(k,r_1,r')\Ical_\ell(k,r_2,r'')\right.
\nonumber\\
&&+\left. P_T(k,z_1,z_2)\Lcal_\ell(k,r_1)\Lcal_\ell(k,r_2)
+\int_0^{r_1}\frac{dr'}{r'}P_T(k,z',z_2)\Ical_\ell(k,r_1,r')\Lcal_\ell(k,r_2)\right.
\nonumber\\
&&\left.+\int_0^{r_2}\frac{dr'}{r'}P_T(k,z_1,z')\Ical_\ell(k,r_2,r')\Lcal_\ell(k,r_1)\right\}\,.
\eea

\section{Scalar power spectra}
Here we want to apply the same procedure used for the tensors also for scalar perturbations. A generic scalar perturbation $A$ can be decomposed as
\begin{equation}
A(z,{\bf n})=\int \frac{d^3{\bf k}}{(2\pi)^3}A(z,{\bf k})\,_0 G_{00}(z,{\bf n},\hat{\bf k})\,.
\end{equation}
With Eqs.~\eqref{eq:basis} and \eqref{eq:rotation}, we have
\begin{equation}
A(z,{\bf n})=4\pi\int \frac{d^3{\bf k}}{(2\pi)^3}A(z,{\bf k})\sum_{\ell=0}^{+\infty}\sum_m(-i)^\ell\,_0f_\ell^{00}(kr)Y_{\ell m}({\bf n};{\bf E})Y^*_{\ell m}(\hat{\bf k};{\bf E})\,.
\label{eq:scalar_dec}
\end{equation}
Eq.~\eqref{eq:scalar_dec} can be readily applied to evaluate the following relevant quantities for lensing theory 
\begin{equation}
\Delta_2 A(z,{\bf n})=-4\pi\int \frac{d^3{\bf k}}{(2\pi)^3}A(z,{\bf k})\sum_{\ell=0}^{+\infty}\sum_m(-i)^\ell\,_0f_\ell^{00}(kr)\ell(\ell+1)Y_{\ell m}({\bf n};{\bf E})Y^*_{\ell m}(\hat{\bf k};{\bf E})\,,
\end{equation}
and
\begin{equation}
\pa^2_{\mp}\pa^2_{\pm}A(z,{\bf n})=\pi\int \frac{d^3{\bf k}}{(2\pi)^3}A(z,{\bf k})\sum_{\ell=0}^{+\infty}\sum_m(-i)^\ell\,_0f_\ell^{00}(kr)\frac{(\ell+2)!}{(\ell-2)!}Y_{\ell m}({\bf n};{\bf E})Y^*_{\ell m}(\hat{\bf k};{\bf E})\,.
\label{eq:scalar_spinor}
\end{equation}

Hence, with Eq.~\eqref{eq:scalar_spinor}, and since $\,_0f_\ell^{00}(x)=j_\ell(x)$, we obtain
\bea
a^{(s)E}_{\ell m}(z)&=& -(-i)^\ell2\pi\sqrt{\frac{\left( \ell+2 \right)!}{\left( \ell-2 \right)!}}
\int \frac{d^3{\bf k}}{(2\pi)^3}\int_0^{r}dr'\frac{r-r'}{r\,r'}\left[\Phi(z',{\bf k})+\Psi(z',{\bf k})\right]\,j_\ell(kr')Y^*_{\ell m}(\hat{\bf k};{\bf E})\,,
\nonumber \\
\eea
whereas $a^{(s)B}_{\ell m}(z)=0$.
This result agrees with the standard literature, see e.g.~\cite{Bartelmann:1999yn}.

In the same way, we also obtain the spectrum for the convergence from scalar perturbations. The harmonic coefficients are
\bea
a^{(s)\kappa}_{\ell m}(z)&=&
(-i)^\ell 4\pi\int \frac{d^3{\bf k}}{(2\pi)^3}\left\{\Psi(z,{\bf k})\,j_\ell(kr)
+\left(1-\frac{1}{\mathcal{H}r}\right)\left[ \Phi(z,{\bf k})\,j_\ell(kr)\right.\right.
\nonumber\\
&&+\int_0^r dr' \left[\Phi'(z',{\bf k})+\Psi'(z',{\bf k})\right]\,j_\ell(kr')
\left.+\int_{0}^{t_s}dt\,\frac{a(t')}{a(t)}\,\Phi(z',{\bf k})\,\pa_rj_\ell(kr) \right]
\nonumber\\
&&-\frac{1}{r}\int_0^r dr'\left[\Phi(z',{\bf k})+\Psi(z',{\bf k})\right]\,j_\ell(kr')
\nonumber\\
&&\left.-\frac{\ell(\ell+1)}{2} \int_0^r dr'\frac{r-r'}{rr'}\left[\Phi(z',{\bf k})+\Psi(z',{\bf k})\right]\,j_\ell(kr')\right\}\,Y^*_{jm}(\hat{\bf k};{\bf E})\,.
\eea
Assuming that $\Phi=\Psi$ as is the case for standard $\La$CDM comology and setting
\bea
\langle \Phi(z_1,{\bf k})\Phi^*(z_2,{\bf q}) \rangle
&=&\langle \Psi(z_1,{\bf k})\Psi^*(z_2,{\bf q}) \rangle
=\langle \Phi(z_1,{\bf k})\Psi^*(z_2,{\bf q}) \rangle
\nonumber\\
&=&(2\pi)^3\delta\left( {\bf k}- {\bf q} \right)P_S(k)T(k,z_1)T(k,z_2)
\,,
\eea
where $P_S(k)$ is the primordial scalar power spectrum and $T(k,z)$ is the tranfer function, we obtain
for the E modes of the shear
\bea
C^{(s)E}_\ell(z_1,z_2)&=&\frac{2}{\pi}\frac{(\ell+2)!}{(\ell-2)!}
\int k^2\,dk\,P_S(k)
\nonumber\\
&&\times\int_0^{r_1}dr'\frac{r_1-r'}{r_1\,r'}T(k,z')\,j_\ell(kr')\int_0^{r_2}dr''\frac{r_2-r''}{r_2\,r''}
T(k,z'')\,j_\ell(kr'')\,,
\label{eq:scalar_shear}
\eea
whereas the angular spectrum of the convergence is
\bea
\label{eq:scalar_conergence}
C^{(s)\kappa}_\ell(z_1,z_2)&=&\frac{2}{\pi}
\int k^2\,dkP_S(k)\left\{T(k,z_1)\,j_\ell(kr_1)
+\left(1-\frac{1}{\mathcal{H}_1r_1}\right)\left[ T(k,z_1)\,j_\ell(kr_1)\right.\right.
\nonumber\\
&&+2\,\int_0^{r_1} dr' \partial_\eta T(k,z')\,j_\ell(kr')
\left.+\int_{0}^{t_1}dt'\,\frac{a(t')}{a(t_1)}\,T(k,z')\,\pa_rj_\ell(kr_1) \right]
\nonumber\\
&&\left.-\frac{2}{r_1}\int_0^{r_1} dr'\,T(k,z')\,j_\ell(kr')
\underbrace{-\ell(\ell+1) \int_0^{r_1} dr'\frac{r_1-r'}{r_1r'}\,T(k,z')\,j_\ell(kr')}\right\}
\nonumber\\
&&\times\left\{T(k,z_2)\,j_\ell(kr_2)
+\left(1-\frac{1}{\mathcal{H}_2r_2}\right)\left[ T(k,z_2)\,j_\ell(kr_2)\right.\right.
\nonumber\\
&&+2\,\int_0^{r_2} dr'' \partial_\eta T(k,z'')\,j_\ell(kr'')
\left.+\int_{0}^{t_2}dt''\,\frac{a(t'')}{a(t_2)}\,T(k,z'')\,\pa_rj_\ell(kr_2) \right]
\\
&&\left.-\frac{2}{r_2}\int_0^{r_2} dr''\,T(k,z'')\,j_\ell(kr'')
\underbrace{-\ell(\ell+1) \int_0^{r_2} dr''\frac{r_2-r''}{r_2r''}\,T(k,z'')\,j_\ell(kr'')}\right\}\,.
\nonumber
\eea
The underbraced terms in Eq.~\eqref{eq:scalar_conergence} are directly sourced by the action of the angular Laplacian on the lensing potential in the convergence, as one sees from the factor $\ell(\ell+1)$. A direct comparison between Eq.~\eqref{eq:scalar_shear} and \eqref{eq:scalar_conergence} shows  that only these terms obey the standard relation
\be
C^{(s)E}_\ell(z_1,z_2)
=\frac{(\ell+2)!}{(\ell-2)!}\frac{C^{(s)\kappa}_\ell(z_1,z_2)}{\ell^2(\ell+1)^2}
=\frac{(\ell+2)(\ell-1)}{\ell(\ell+1)}\,C^{(s)\kappa}_\ell(z_1,z_2) \,.
\ee

For large $\ell$, one can simplify \eqref{eq:scalar_conergence}. But since here we are interested in the difference between the $\ka$ and $\ga_E$ power spectra which are most relevant at low $\ell$, we do not use the Limber approximation here.

Let us finally remark that the structure of Eq.~\eqref{eq:scalar_conergence} agrees with the one found in \cite{Fanizza:2021tuh}, where the $C_\ell$'s of the distance-redshift relation for scalar perturbations have been studied. At linear level,  distance perturbations are actually equal to  the convergence (up to a sign). As already pointed out in \cite{Fanizza:2021tuh}, the monopole of the convergence is not affected at all by lensing corrections.
Let us also mention that making use of momentum conservation (the relativistic Euler equation for pressure-less matter yields $a(t_1)V(t_1) =\int_0^{t_1} dt a\Psi$), the time integrals in \eqref{eq:scalar_conergence} can actually be replaced by the matter velocity field. We have used this for the numerical evaluation of the $\ka$-spectrum in Section~\ref{s:results}.

\section{Evolution of shear E and B modes}\label{a:evolution}

Let us start from the evolution equation for the Jacobi map (see \cite{Fleury:2013sna,Fanizza:2013doa} for details)
\be
\frac{d^2 J_{AB}}{d\lambda^2}=R_{AC} J_{CB}\,,
\label{eq:Jev}
\ee
where $R_{AC}\equiv R_{\alpha\beta\nu\mu}k^\alpha k^\nu s^\beta_A s^\mu_C$ is a projection of the Riemann tensor on the Sachs basis and $k^\mu$. We want to derive the evolution equation for the lensing shear. To this end, we project Eq.~\eqref{eq:Jev} on the Pauli matrices $\sigma_{3/1}$. With the definition given in Eqs.~\eqref{eq:39}, we have for the lhs of Eq.~\eqref{eq:Jev}
\be
\frac{d^2 \left(J_{AB}\sigma^{AB}_3\right)}{d\lambda^2}
=2\frac{d^2\left(\bar{d}\ga_1\right)}{d\lambda^2}
=2\bar{d}\,\frac{d^2 \ga_1}{d\lambda^2}
+4\frac{d\,\bar{d}}{d\lambda}\frac{d\ga_1}{d\lambda}
+2\frac{d^2\bar{d}}{d\lambda^2}\ga_1\,,
\ee
and the same for $\ga_2$. On the other hand $R_{AC}$ can be decomposed in terms of the Ricci focusing $\Phi_{00}$ and the Weyl projection $\Psi_0$ (see \cite{Fanizza:2014baa}) as
\be
R_{AC}=\Phi_{00}\,\delta_{AC}
+\text{Re}\Psi_0\,(\sigma_3)_{AC}
+\text{Im}\Psi_0\,(\sigma_1)_{AC}\,,
\ee
where $\Phi_{00}\equiv -\frac{1}{2}R_{\mu\nu}k^\mu k^\nu$ and $\Psi_0\equiv \frac{1}{2}C_{\alpha\beta\mu\nu}k^\alpha k^\mu \Sigma^\beta \Sigma^\nu$ are respectively sourced by Ricci tensor $R_{\mu\nu}$ and the Weyl tensor $C_{\alpha\beta\mu\nu}$, being $\Sigma^\mu\equiv s^\mu_1+i s^\mu_2$. When projected on the Pauli matrices, using that $\omega =0$, the rhs of Eq.~\eqref{eq:Jev} becomes
\bea
R_{AC} J_{CB}\sigma_3^{AB}
&=&2\bar{d}\left[ (1-\kappa)\text{Re}\Psi_0+\ga_1\,\Phi_{00}  \right]\,,
\nonumber\\
R_{AC} J_{CB}\sigma_1^{AB}
&=&2\bar{d}\left[ (1-\kappa)\text{Im}\Psi_0+\ga_2\,\Phi_{00} \right]\,.
\eea
Defining $\dot{\,}\equiv d/d\lambda$, we find that the evolution equations for the shear components become
\bea
\ddot{\ga}_1
+2\frac{\dot{\bar{d}}}{\bar{d}}\,\dot{\ga}_1
+\frac{\ddot{\bar{d}}}{\bar{d}}\,\ga_1
&=&(1-\kappa)\text{Re}\Psi_0+\ga_1\,\Phi_{00}\,,
\nonumber\\
\ddot{\ga}_2
+2\frac{\dot{\bar{d}}}{\bar{d}}\,\dot{\ga}_2
+\frac{\ddot{\bar{d}}}{\bar{d}}\,\ga_2
&=&(1-\kappa)\text{Im}\Psi_0+\ga_2\,\Phi_{00}\,,
\eea
leading to
\bea
\ddot{\ga}_+
+2\frac{\dot{\bar{d}}}{\bar{d}}\,\dot{\ga}_+
+\frac{\ddot{\bar{d}}}{\bar{d}}\,\ga_+
&=&(1-\kappa)\Psi_0+\ga_+\,\Phi_{00}\,,
\nonumber\\
\ddot{\ga}_-
+2\frac{\dot{\bar{d}}}{\bar{d}}\,\dot{\ga}_-
+\frac{\ddot{\bar{d}}}{\bar{d}}\,\ga_-
&=&(1-\kappa)\overline{\Psi}_0+\ga_-\,\Phi_{00}\,.
\label{eq:shear_evolution}
\eea

So far these equations are exact, in the absence of rotation. At linear order, we have that $\ga_\pm$ is a purely linear quantity, so that we can take the background value of $\Phi_{00}$, which is independent of direction. On the other hand, $\Psi_0$ vanishes on the background, hence for $\kappa$ we must insert the background value which is $0$. Finally, at linear order $d/d\lambda$ involves only time and radial derivatives, hence Eqs.~\eqref{eq:shear_evolution} can be readily converted to equations for the $a^\pm_{\ell m}$'s as
\be
\ddot{a}^\pm_{\ell m}
+2\frac{\dot{\bar{d}}}{\bar{d}}\,\dot{a}^\pm_{\ell m}
+\frac{\ddot{\bar{d}}}{\bar{d}}\,a^\pm_{\ell m}
=a^{\Psi_0\,\pm}_{\ell m}+\Phi_{00}\,a^\pm_{\ell m}\,,
\ee
where we have decomposed also $\Psi_0$ in terms of the $\,_{\pm 2}Y_{\ell m}$. We then obtain
\be
\ddot{a}^{E/B}_{\ell m}
+2\frac{\dot{\bar{d}}}{\bar{d}}\,\dot{a}^{E/B}_{\ell m}
+\frac{\ddot{\bar{d}}}{\bar{d}}\,a^{E/B}_{\ell m}
=a^{\Psi_0\,E/B}_{\ell m}+\Phi_{00}\,a^{E/B}_{\ell m}\,.
\ee

But $a^{\Psi_0\,E/B}_{\ell m}$ are simply the harmonic coefficients of the Weyl tensor's $E$- and $B$-modes.
The $E$- and $B$-part of the Weyl tensor from gravitational waves is, see~\cite{durrer_2020}, Appendix~A3.3.4,
\bea
E_{ij} &=& \frac{1}{2}\left(\dd_t^2 +\De\right)H_{ij} \simeq -k^2H_{ij}\\
B_{ij} &=&-\frac{1}{2}\ep_{n m i}\dd_t\left(\dd_mH_{jn}-\dd_nH_{jm}\right) \simeq
-kk_mH_{jn}\ep_{nmi} \,.
\eea
Here the $\simeq$ sign is valid for a fixed $k$-mode inside the horizon when $\pa_tH \simeq ikH$ and $\pa_mH\simeq ik_mH$. For this it is also important that the gravitational wave obeys the free wave equation and is not sourced by some tensor anisotropic stress. Such a source term would modify the relation. Let us consider the concrete example of an orthonormal system $\hat\bk,\boe^{(1)},\boe^{(2)}$, where $\bk$ is the wave vector of our gravitational wave, and 
$\boe^{(2)}=\hat\bk\times\boe^{(1)}$. We can orient this basis such that in Fourier space
$$ H_{ij} = H(\bk,t)\left(\begin{array}{ccc}0 & 1 & 0 \\ 1 & 0 & 0 \\ 0 & 0& 0\end{array}\right)
= H(\bk,t)\left(\boe^{(1)}_i\boe^{(2)}_j +\boe^{(2)}_i\boe^{(1)}_j\right)
\,.$$
Inserting this above we obtain
\bea
E_{ij} &\simeq& k^2H(\bk,t)\left(\boe^{(1)}_i\boe^{(2)}_j +\boe^{(2)}_i\boe^{(1)}_j\right)\,,\\
B_{ij} &\simeq&  k^2H(\bk,t)\left(\boe^{(1)}_i\boe^{(1)}_j -\boe^{(2)}_i\boe^{(2)}_j\right)\,.
\eea
While these components are uncorrelated, they have both the same power spectrum given by $2k^4P_T(k)$. Therefore free gravitational waves generate shear $E$- and $B$-modes with the same power spectra on subhorizon scales, i.e., for $k>\HH(z)$.

\section{Spin-weighted spherical harmonics}
\label{app:swsh}
In order for the paper to be as self-contained as possible, 
 we provide here some useful relations of spin-weighted spherical harmonics and the spin raising and lowering operators which are extensively used in this work. This appendix is not meant to give an exhaustive discussion of the subject. It rather provides the reader with some basic useful formulas needed to obtain our results. Anyone interested in a more detailed discussion about the topic is referred to \cite{Hu:2000ee,Bernardeau:2009bm,Schmidt:2012ne,Diss-Seibert}.

Let us start by considering a function $f(\bf n)$ with spin $s$. We define the spin raising and lowering operators respectively as
\bea
\ds f({\bf n}) &\equiv& -\sin^s\theta\left[ \pa_\theta+\frac{i}{\sin\theta}\pa_\phi \right]\left[\sin^{-s}\theta\,f({\bf n})\right]
=-\left(\pa_\theta+\frac{i}{\sin\theta}\pa_\phi\right)f({\bf n})
+s\cot\theta f({\bf n})
\nonumber\\
\bds f({\bf n}) &\equiv& -\sin^{-s}\theta\left[ \pa_\theta-\frac{i}{\sin\theta}\pa_\phi \right]\left[\sin^s\theta\,f({\bf n})\right]
=-\left(\pa_\theta-\frac{i}{\sin\theta}\pa_\phi\right)f({\bf n})
-s\cot\theta f({\bf n})\,.
\nonumber\\
\label{eq:spin_operators}
\eea
Here $(\theta,\phi)$ are the usual polar angles of the direction {\bf n}.
The effect of the operator $\ds$ ($\bds$) on the function $f$ is to raise (lower) its spin, such that $\ds f$ ($\bds f$) has spin $s+1$ ($s-1$). For a spin zero function $\bds f=0$. In this way, the spin-weighted spherical harmonics $_sY_{\ell m}$ are obtained from the spin zero spherical harmonics $Y_{\ell m}$ by acting with $\ds$ and $\bds$ respectively
\be
\,_sY_{\ell m}({\bf n};{\bf E})=
\sqrt{\frac{\left(\ell-|s|\right)!}{\left(\ell+|s|\right)!}}
\begin{cases}
\ds^s Y_{\ell m}({\bf n};{\bf E})\qquad&,\qquad s\ge 0\\
\left(-1\right)^s \bds^{|s|}Y_{\ell m}({\bf n};{\bf E})\qquad&,\qquad s<0 \,.
\end{cases}
\ee
The prefactor is needed to ensure that they remain normalized, i.e.
\be
\int d\Omega_{\bf n}
\,_sY_{\ell m}({\bf n};{\bf E})
\,_sY^*_{\ell' m'}({\bf n};{\bf E})
=\delta_{\ell\ell'}\delta_{mm'}\,.
\label{eq:D3}
\ee
Our notation is such that the pair ({\bf n};{\bf E}) indicates that the angles of the direction ${\bf n}$ are taken with respect to the reference frame where ${\bf E}$ is the $z$-direction. Hence, the angular derivatives in the definition of the spin operators and the dependence on ${\bf n}$ in Eq.~\eqref{eq:spin_operators} are  referring to the same basis.

With the use of Eq.~\eqref{eq:38}, we can write Eqs.~\eqref{eq:spin_operators} as
\bea
\ds f({\bf n}) &=& \left(-\sqrt{2}\,\hat{s}^a_+\pa_a + s\cot\theta \right) f({\bf n})= -\sqrt{2}\,\pa_+ f({\bf n})\,,
\nonumber\\
\bds f({\bf n}) &=& \left(-\sqrt{2}\,\hat{s}^a_-\pa_a - s\cot\theta \right) f({\bf n})= -\sqrt{2}\,\pa_- f({\bf n})\,,
\eea
where $\pa_\pm\equiv\hat{s}^a_\pm\pa_a\mp s\frac{\cot\theta}{\sqrt{2}}$. A useful relation for us, concerning the derivatives of the basis, is (see \cite{Bernardeau:2009bm})
\be
\sqrt{2}\,r\,e^i_\pm \pa_i e^j_\pm = \cot\theta\,e^j_\pm\,.
\ee
This relation can be rewritten in terms of our basis $\hat{s}^a_\pm$, again by making use of Eq.~\eqref{eq:38}, as follows
\be
\hat{s}^a_\pm \nabla_a\hat{s}^c_\pm=
\frac{\cot\theta}{\sqrt{2}}\,\hat{s}^c_\pm\,.
\label{eq:D6}
\ee

As a final remark, we note that Eq.~\eqref{eq:D6} allows us to explicitly deal with double projections on $\hat{s}^a_\pm$ in Eq.~\eqref{eq:311}. Indeed, for a generic scalar field $A({\bf n})$, we have that
\bea
\hat{s}^a_+\hat{s}^b_+\nabla_a\pa_{b}A({\bf n})
&=&\hat{s}^a_+\pa_a\left(\hat{s}^b_+\pa_bA({\bf n})\right)
-\hat{s}^a_+\nabla_a\left(\hat{s}^b_+\right)\pa_bA({\bf n})
\nonumber\\
&=&-\frac{1}{\sqrt{2}}\hat{s}^a_+\pa_a\ds A({\bf n})
-\frac{\cot\theta}{\sqrt{2}}\,\hat{s}^c_+\pa_cA({\bf n})
\nonumber\\
&=&-\frac{1}{\sqrt{2}}\hat{s}^a_+\pa_a\ds A({\bf n})
+\frac{\cot\theta}{2}\,\ds A({\bf n})
=\frac{1}{2}\ds^2 A({\bf n})\,,
\label{eq:D7}
\eea
and the same holds also for $\hat{s}^a_-\hat{s}^b_-\nabla_a\pa_{b}A({\bf n})=\frac{1}{2}\bds^2 A({\bf n})$.

\bibliographystyle{JHEP}
\bibliography{biblio_shear}

\end{document}